\documentclass[aps,nofootinbib]{revtex4}

\usepackage{amsmath}
\usepackage{amssymb}
\usepackage{bbm}
\usepackage{mathtools}
\usepackage[normalem]{ulem}
\usepackage{dsfont}
\usepackage{mathrsfs}
\usepackage[makeroom]{cancel}
\usepackage{graphicx}
\usepackage{bm}

\def\rhat{{\hat {\boldsymbol r}}}
\def\khat{{\hat {\boldsymbol k}}}

\renewcommand{\sinh}{\operatorname{sh}}
\renewcommand{\cosh}{\operatorname{ch}}

\renewcommand{\coth}{\operatorname{cth}}

\DeclareMathOperator\sign{sgn}
\DeclareMathOperator\Si{Si}

\def\0{\boldsymbol{0}}
\def\RR{\mathbbm{R}}
\def\ZZ{\mathbbm{Z}}
\def\CC{\mathbbm{C}}
\def\NN{\mathbbm{N}}

\def\BD#1{{\boldsymbol{{#1}}}}

\def\Lag{{\cal L}}
\def\ii{\mathrm{i}}

\def\bnabla{{\boldsymbol\nabla}}

\def\VAREPS#1{\varepsilon_{#1}}
\def\OM#1{\omega_{#1}}

\def\B#1{\left(#1\right)}
\def\BB#1{\left[#1\right]}
\def\BBB#1{\left|#1\right|}

\def\la{\langle}
\def\ra{\rangle}
\def\lara#1{\la#1\ra}

\def\be{\begin{equation}}
\def\ee{\end{equation}}

\def\hc{\text{h.c.}}
\def\cc{\text{c.c.}}

\def\for{\ \text{for} \ }

\usepackage[ddmmyyyy]{datetime}
\makeatletter
\def\Dated@name{}
\makeatother

\def\pref{0.6}
\def\CL{\text{cl}}

\begin{document}
\title{Non-equilibrium dynamics of long-range field configurations in the Proca theory
and the counterexample to the law of  periodic charge oscillations  }
\author{Bogdan Damski}
\affiliation{Jagiellonian University, 
Faculty of Physics, Astronomy and Applied Computer Science,
{\L}ojasiewicza 11, 30-348 Krak\'ow, Poland}
\begin{abstract}
Long-range field configurations exist in the Proca theory and 
their non-equilibrium evolution is of interest in this work.
General arguments suggest that a charge can be assigned to them and that its
evolution is governed by the law of periodic charge oscillations.
We discuss an elegant analytically-solvable example of a field 
configuration in the Proca theory respecting such a law. 
We also identify a weak point in the 
aforementioned  general arguments, construct the counterexample
to the law of periodic charge oscillations in the Proca theory, and comprehensively discuss it.
The Gibbs-Wilbraham phenomenon is discussed in the course of  these studies.
\end{abstract}
\maketitle

\section{Introduction}
\label{Introduction_sec}
 
We explore  in this work  
 non-equilibrium dynamics of the  
Proca theory, whose Lagrangian density 
is given by the following formula
\be
\Lag=-\frac{1}{4} F^\CL_{\mu\nu}F_\CL^{\mu\nu}
+\frac{m^2}{2}V^\CL_\mu V_\CL^\mu,
\label{SPr}
\ee
where $F^{\mu\nu}_\CL=\partial^\mu V_\CL^\nu-\partial^\nu V_\CL^\mu$, 
 $V_\CL$ is the vector field,
 $m>0$ is the mass of the vector 
boson, and the Heaviside-Lorentz system of units
with $\hbar=c=1$ is assumed.\footnote{The remaining  conventions
are the following. 
We use the metric tensor 
$\text{diag}(1,-1,-1,-1)$ and assume that 
Greek (Latin) indices of tensors 
take values $0,1,2,3$ ($1,2,3$). 
We use the Einstein summation convention.
$3$-vectors are written in bold, 
e.g. $V=(V^\mu)=(V^0,\BD{V})$,
the hermitian (complex)
conjugation is denoted as $\hc$ ($\cc$),
$x^+$ ($x^-$) denotes
the quantity that is infinitesimally larger (smaller)
than $x$, and $\hat{\BD{x}}=\BD{x}/|\BD{x}|$.
}
The subscript $\CL$ is introduced to distinguish classical
expressions from their quantum counterparts  discussed  in 
the majority of this work.

Despite its simplicity,  Proca theory (\ref{SPr}) serves as a useful platform for
illustrating various field-theoretic concepts.
Moreover, it captures certain properties of massive vector bosons
such as  $\rho$ and $\omega$ mesons \cite{Greiner,ColemanBook,Weinberg}. 
However, these particles are short-lived, which is problematic  in the context of our study.
The Proca theory is traditionally regarded as an extension of Maxwell
electrodynamics \cite{Nieto_RMP1971,Gillies2005,Nieto_RMP2010}, 
the one in which a photon is treated as a massive particle. 
Modified versions of the Proca theory are extensively 
discussed in a cosmological context \cite{HeisenbergPhysRep2019}.

To explain the basic idea behind this research, we write Proca 
field  equations as
\begin{align}
\label{EL}
&\partial_\mu F^{\mu\nu}_\CL={\cal J}_\CL^\nu,\\
&{\cal J}^\nu_\CL=-m^2 V_\CL^\nu,
\end{align}
where (\ref{EL})  implies
\be
\partial_\nu {\cal J}_\CL^\nu=0
\label{curJ}
\ee
expressing the fact that  ${\cal J}_\CL$ is a locally
conserved $4$-current (an internal  $4$-current, so to
speak). As the timelike component of a locally conserved
$4$-current 
is generally considered  as a charge density, we 
introduce the charge
\be
Q_\CL(t,\RR^3) = \int d^3r {\cal J}^0_\CL(t,\BD{r}),
\label{Qclass}
\ee
where $\RR^3$ refers to the integration domain
and we observe that such a definition 
mirrors the Maxwell theory counterpart of the
studied quantity due to  
${\cal J}^0_\CL=\partial_i F^{i0}_\CL$.

Then, we note that it is generally known that  local 
conservation of a $4$-current 
does not imply  conservation of the corresponding 
charge. Field configurations in the Proca theory, where 
the charge is  time dependent or 
poorly defined via the above integral,
will be  studied in this  work.

Finally, we observe  that one may  infer 
from (\ref{curJ}) that 
the vector field of the Proca theory is Lorenz gauge fixed
(the gauge invariant version of the
Proca theory also exists;
see e.g.  \cite{Guralnik1968,Pimentel2015,BDGaugeinv1}).
With a little more effort  \cite{Nieto_RMP1971}, 
one may actually argue that 
such a vector field 
is in principle observable (see e.g.
\cite{LakesPRL1998} reporting experimental 
efforts towards its measurement).
We mention in passing that 
surprisingly diverse experimental approaches,
targeting differences between the Proca and
Maxwell theories, are comprehensively reviewed in
\cite{Nieto_RMP1971,Gillies2005,Nieto_RMP2010}.

The outline  of this paper is the following. 
Sec. \ref{Quant_sec} presents basic facts about the quantum Proca theory
as well as the terminology used throughout the paper. 
Sec. \ref{Charged_sec} discusses the construction of charged states in the
Proca theory 
and offers a critical presentation of the derivation of the law of periodic charge oscillations
in such  a theory. 
A detailed analytical illustration of periodic charge oscillations 
in a classical field configuration within the Proca theory is provided in Sec. \ref{No_sec}.
Secs. \ref{Sharp_sec} and \ref{Large_sec} focus on a quantum field configuration 
in the Proca theory
constructed to serve as a counterexample to the law of periodic charge oscillations.
Specifically, Sec. \ref{Sharp_sec} provides rigorous results regarding the 
regularity and long-distance properties of the studied field configuration.
Sec. \ref{Large_sec} extends this discussion to all distances from the center
of the field configuration and establishes the presence of the Gibbs-Wilbraham 
phenomenon in the problem under study. 
The results in  Sec. \ref{Large_sec} are obtained using an appropriate approximation that 
relies on the findings  from Sec. \ref{No_sec}.
A summary of our work  is given in Sec. \ref{Summary_sec}.

\section{Quantization of Proca theory }
\label{Quant_sec}

Quantization of the vector field of the 
Proca theory leads to  \cite{Greiner}
\begin{subequations}
\be
V^\mu(t,\BD{r})= \int
\frac{d^3k}{(2\pi)^{3/2}}\frac{1}{\sqrt{2\VAREPS{k}}}
  \sum_{\sigma=1}^3
\eta^\mu(\BD{k},\sigma) a_{\BD{k}\sigma}\exp(-\ii\VAREPS{k}t + \ii\BD{k}\cdot\BD{r}) + \hc,
\ee
where   annihilation and  creation  operators
satisfy   ($\sigma,\sigma'=1,2,3$)
\be
 [a_{\BD{k}\sigma},a^\dag_{\BD{k}'\sigma'}]=\delta_{\sigma\sigma'}\delta(\BD{k}-\BD{k}'),
 \
[a_{\BD{k}\sigma},a_{\BD{k}'\sigma'}]=0,
\ee
transverse polarization $4$-vectors  obey ($i,j=1,2$)
\begin{align}
&\eta(\BD{k},i)=\B{0,\boldsymbol{\eta}(\BD{k},i)}, \ \boldsymbol{\eta}(\BD{k},i)\in\RR^3, \\
& \boldsymbol{\eta}(\BD{k},i)\cdot\BD{k}=0, \
 \boldsymbol{\eta}(\BD{k},i)\cdot\boldsymbol{\eta}(\BD{k},j)=\delta_{ij},
\end{align}
the longitudinal polarization $4$-vector reads
\be
\eta(\BD{k},3)=\B{\frac{\OM{k}}{ m},\frac{\VAREPS{k}}{ m}\khat},
\ee
\label{Vvecoperator}%
\end{subequations}
  operators 
$a_{\BD{k}\sigma}$
annihilate 
the vacuum state $|0\ra$,
$\OM{k}=|\BD{k}|$, and 
$\VAREPS{k}=\sqrt{m^2+\OM{k}^2}$.

Next, we introduce the electric field operator 
\be
\label{piE}%
\BD{E}(t,\BD{r})= -\bnabla V^0(t,\BD{r})-\partial_t\BD{V}(t,\BD{r})=
\ii\int \frac{d^3k}{(2\pi)^{3/2}}
\sqrt{\frac{\VAREPS{k}}{2}}
\sum_{\sigma=1}^3
\B{1-\frac{\OM{k}^2}{\VAREPS{k}^2}\delta_{\sigma3}}
 \boldsymbol{\eta}(\BD{k},\sigma) 
a_{\BD{k}\sigma}\exp(-\ii\VAREPS{k}t + \ii\BD{k}\cdot\BD{r}) + \hc,
\ee
where $\bnabla=(\partial/\partial r^i)$ and $\partial_t=\partial/\partial t$.
The above-proposed name of such an operator comes from the 
fact that (\ref{piE})   is  defined
in terms of  the vector field just  as the electric
field of the Maxwell theory.
Moreover, we will refer to  the
expectation value of the electric field operator, say 
$\lara{\BD{E}(t,\BD{r})}$, 
as the electric field for the sake of brevity.
By the same token, 
\begin{align}
\label{Bb}
&\lara{\BD{B}(t,\BD{r})}=\lara{\bnabla\times\BD{V}(t,\BD{r})},\\
\label{j00}
&\lara{{\cal J}^0(t,\BD{r})}=\lara{\bnabla\cdot\BD{E}(t,\BD{r})},\\
&\lara{\BD{\cal J}(t,\BD{r})}=\lara{-\partial_t\BD{E}(t,\BD{r})},
\label{VpartT}
\end{align}
will be referred to as the magnetic field, the charge density, and the $3$-current.
While (\ref{Bb}) and (\ref{j00}) have an obvious origin,
(\ref{VpartT}) follows from
the ``expectation value'' of Proca field equations combined with (\ref{B0}).

When using such  terminology,  one should keep in mind that 
we work here with massive vector  bosons and their theory 
is fundamentally different from the  
Maxwell theory \cite{Greiner,ColemanBook,Weinberg,Nieto_RMP1971,Gillies2005,Nieto_RMP2010}.
Such a remark is  important  because 
the states of interest in this work  
will be entirely built out of longitudinally-polarized 
vector  bosons having no counterpart in the Maxwell theory.
Moreover, one should be aware  that  the term charge 
refers to the quantity characterizing  field configurations 
(not  properties of electrically charged particles).
We mention in passing that 
the  magnetic field operator is not  discussed  in our work because 
it does not have a longitudinally-polarized  component.
This implies that in the studied states   
\be
\lara{\BD{B}(t,\BD{r})}=\0.
\label{B0}
\ee

Finally, the Hamiltonian of the Proca
theory is given by \cite{Greiner}
\be
H=\int d^3k\, \VAREPS{k}\sum_{\sigma=1}^3
a_{\BD{k}\sigma}^\dag a_{\BD{k}\sigma}.
\label{HProca}
\ee
Its time independence implies  conservation of the energy 
of the studied field configurations.

\section{Charged states}
\label{Charged_sec}
We define charged states as the states in which the 
expectation value of the 
electric field operator, i.e. the electric field,  exhibits  
the inverse-square  asymptotic decay,
which   could be modulated by 
oscillatory terms.  Such states encode long-range 
field configurations. 
Next, we briefly summarize the way charged states 
were introduced in \cite{BDPeriodic1}. Namely,
without going into too many details, 
they were defined as the properly-weighted superpositions 
of the vacuum state $|0\ra$ and the state 
$\chi|0\ra$,
where 
\be
\chi=
\frac{\ii q}{m} \int\frac{d^3k}{(2\pi)^{3/2}}
f(\OM{k})
\sqrt{\frac{\VAREPS{k}}{2\OM{k}^2}}
a_{\BD{k}3} + \hc
\ee
with  $\RR\ni q\neq0$ and  $f$ being a dimensionless real function
normalized such that $f(0)=1$ (the same is also assumed 
in this work).
For $f$'s studied in \cite{BDPeriodic1}, 
such states represent  field configurations 
 centered at $\0$ in  space 
and having the charge  $\propto q$.

In this  work, we define charged states as
\be
|\psi\ra= \exp\B{-\ii\chi}|0\ra,
\label{psiCoh}
\ee
which leads to the following observations.

First, the state $|\psi\ra$ is, by construction, normalized to unity. 
Interestingly, normalizability of  the charged states 
studied in 
\cite{BDPeriodic1}
required the consideration of $f(\OM{k})$ vanishing sufficiently fast
for $\OM{k}\to\infty$.

Second, we introduce an operator $O$ that is  
linear in
 creation and annihilation operators
and has zero vacuum expectation value 
($O=\BD{E}(t,\BD{r})$, 
$a_{\BD{k}3}$, etc.).
For such an operator 
\be
\lara{O}=-\ii  [O, \chi],
\label{Ok}
\ee
where $\lara{\cdots}$ denotes $\lara{\psi|\cdots|\psi}$
till the end of this work and $ [O, \chi]\in\CC$. 
This result has been obtained   using  
\be
[O,\exp(-\ii\chi)]=-\ii[O,\chi]\exp(-\ii\chi),
\label{Okom}
\ee
which holds   in the discussed circumstances.

Third, the state   $|\psi\ra$ represents a coherent state (see 
e.g. \cite{PandA} for a textbook  
discussion of coherent states).  Namely,
\be
a_{\BD{k}3}|\psi\ra = 
-\frac{q}{m} \frac{f(\OM{k})}{(2\pi)^{3/2}}
\sqrt{\frac{\VAREPS{k}}{2\OM{k}^2}}|\psi\ra,
\label{aCoH}
\ee
which can be proved by means of   (\ref{Okom}).
Such a result 
allows for an easy computation of 
the energy of the field configuration described by
(\ref{psiCoh}) 
\be
\lara{H}= \frac{q^2}{4\pi^2 m^2}\int_0^\infty d\OM{k} 
 \VAREPS{k}^2 f^2(\OM{k}).
 \label{Hf}
\ee

Fourth, there are infinitely  many states $|\Psi\ra$ such that 
\be
\lara{\Psi|O|\Psi}=\lara{O}
\label{OO1}
\ee
for all operators  $O$ having  the properties specified above (\ref{Ok}).
In the context of our work, this implies that the  electric field,
magnetic field,  etc. do 
not uniquely determine the quantum state of the studied field configuration.
To prove the above statement, one may  consider 
\be
|\Psi\ra= \exp(-\ii\chi)|\alpha\ra, 
\ee
where   
$\lara{\alpha|\alpha}=1$ and $\lara{\alpha|O|\alpha}=0$,
then use (\ref{Okom})  to argue that it leads to   (\ref{OO1}), 
and finally note that there are infinitely many states $|\alpha\ra$
satisfying the above conditions (e.g. all 
    properly-normalized  superpositions
of  Fock states having an  even total  occupation of momentum 
modes; the simplest case in point  is given  by 
$|\alpha\ra=\int d^3k\, d^3k' f(\BD{k},\BD{k}') a^\dag_{\BD{k}\sigma}
a^\dag_{\BD{k}'\sigma'}|0\ra$, where
$\sigma,\sigma'=1,2,3$ and 
$f(\BD{k},\BD{k}')\in\CC$ is chosen 
such that  $\lara{\alpha|\alpha}=1$).
Quite remarkably, nothing of this kind exists on the classical level,
where knowledge of the electric and  magnetic field
uniquely identifies the  field configuration.
 It should be also said 
that expectation  values of operators, which  are not linear in 
the creation and annihilation operators, will generally 
depend on whether they are computed in the state
$|\psi\ra$ or  $|\Psi\ra$. 
For example,  one may easily  verify that 
\be
\lara{\Psi|H|\Psi} = \lara{\alpha|H|\alpha} + \lara{H}
\label{HH1}
\ee
for $\lara{\alpha|a_{\BD{k}3}|\alpha}=0$.

Fifth, $\exp(-\ii\chi)$ can be seen as an operator 
{\it additively} imprinting 
a longitudinally-polarized  field configuration on top of any other 
field configuration. To explain this remark, 
we  repeat the above calculations 
without assuming  $\lara{\alpha|O|\alpha}=0$,
which results in  
\be
\lara{\Psi|O|\Psi}=\lara{\alpha|O|\alpha} + \lara{O},
\label{OO2}
\ee
where one may e.g. substitute $\BD{E}(t,\BD{r})$  
for $O$ to see  physical implications of the 
above formula.
For the sake of completeness, we also note that 
without assuming $\lara{\alpha|a_{\BD{k}3}|\alpha}=0$,
we arrive at   
\be
\lara{\Psi|H|\Psi} = \lara{\alpha|H|\alpha} + \lara{H}
 -\frac{q}{m} \int \frac{d^3k}{(2\pi)^{3/2}} f(\OM{k})
\sqrt{\frac{\VAREPS{k}^3}{2\OM{k}^2}} 
[\lara{\alpha|a_{\BD{k}3}|\alpha}+\cc],
\label{HH2}
\ee
which, unlike (\ref{HH1}) and (\ref{OO2}), 
does not seem to have  a
straightforward physical interpretation.

To discuss the charge associated with
the field configuration described by 
wave-function (\ref{psiCoh}), we 
introduce the charge enclosed in the area $\cal V$ via the following formula
\be
\lara{Q(t,{\cal V})}=\int_{\cal V} d^3r  \lara{\bnabla\cdot\BD{E}(t,\BD{r})},
\label{QV}
\ee
where  we
  substitute $r<R$ ($r>R$) for  $\cal V$ when we 
discuss the charge inside (outside) the ball of the radius $R$ [such a ball,
the spherical shell referred to  below, 
and all  field configurations studied in this work are centered at $\0$
in  space].
With the help of electric field operator
(\ref{piE}), this can be rewritten to the form 
\be
\lara{Q(t,{\cal V})}=-m \int_{\cal V}d^3r\int\frac{d^3k}{(2\pi)^{3/2}}
\frac{\OM{k}}{\sqrt{2\VAREPS{k}}} \lara{a_{\BD{k}3}}
\exp(-\ii\VAREPS{k}t+\ii\BD{k}\cdot\BD{r})
+\cc
\label{QtV}
\ee

If we now assume that $\lara{Q(t,r<R)}$ converges to a well-defined   
value in the 
limit of $R\to\infty$, we can simplify the above expression
by replacing $\int_{\cal V}d^3r$ with $\lim_{\epsilon\to0^+}\int d^3r
\exp(-\epsilon |\BD{r}|^2)$,  commuting $d^3r$ and $d^3k$ integrals,
and finally doing the $d^3r$ integration. In the end, we arrive at 
\begin{align}
&\lara{Q(t, \RR^3)}=\lim_{\epsilon\to0^+} \lara{Q_\epsilon(t,\RR^3)},\\
&\lara{Q_\epsilon(t,\RR^3)}= -\sqrt{\frac{ m}{2}}(2\pi)^{3/2}\int
d^3k\,\OM{k}\delta_\epsilon(\BD{k})  \lara{a_{\BD{k}3}}\exp(-\ii m t)+\cc,\\
&\delta_\epsilon(\BD{k})=  \frac{1}{(2\sqrt{\pi\epsilon})^3}
\exp\B{-\frac{\OM{k}^2}{4\epsilon}},
\end{align}
where $\delta_\epsilon(\BD{k})$ is  a nascent delta function.
It is then easy to note that such  $\lara{ Q(t, \RR^3)}$ satisfies the harmonic 
oscillator equation,
\be
\frac{d^2}{dt^2}\lara{Q(t, \RR^3)}=-m^2 \lara{Q(t, \RR^3)},
\label{d2Q}
\ee
which we term as the  {\it law of periodic charge  oscillations}.
Equation of such a sort
was  merely stated in seminal paper 
\cite{GHK1964} and review \cite{Guralnik1968}
in the context of symmetry breaking studies.
The discussion of  periodic charge
oscillations in the Proca theory can be found in \cite{BDPeriodic1,BDNPB2024ref}.
Similar phenomenon was mentioned 
in the context of spatially unbounded superconductors in \cite{Hertzberg2020}.
The periodic oscillations of dipole moments 
in the Proca theory were described in \cite{BDDipole1}.

For state (\ref{psiCoh}), we find 
\be
\lara{ Q(t,\RR^3)}=q\cos(mt),
\label{Qtt}
\ee
which satisfies  (\ref{d2Q}) and 
reveals  the meaning of the parameter $q$.
The problem with the above-presented  reasoning, however, 
is that it assumes existence of the 
$R\to\infty$ limit of $\lara{Q(t,r<R)}$, which should not be taken for granted. 
In fact, we missed such a point in our earlier studies (see e.g. 
\cite{BDPeriodic1}). The concrete example, where 
 $\lim_{R\to\infty}\lara{Q(t,r<R)}$ is undefined, and as such 
(\ref{d2Q}) and (\ref{Qtt}) are meaningless, will be 
 discussed in Secs. \ref{Sharp_sec} and \ref{Large_sec}.

Finally, we note that unless stated otherwise, 
\be
t>0
\ee 
is assumed in Secs. \ref{No_sec}--\ref{Large_sec}
for the sake of convenience.
The straightforward extension of our findings to an arbitrary time is discussed 
in Sec. \ref{Summary_sec}.

\section{No cutoff in momentum space}
\label{No_sec}

A classical field configuration, exhibiting  periodic charge oscillations, 
is  constructed and analytically analyzed in this section. 
We demonstrate  that a spherical shock-wave 
appears  in the discussed  setting and examine  the associated electric field 
and related quantities to illustrate this  feature.
In particular, we provide a detailed discussion of the charge localized 
in different regions of space 
(on the shock-wave front as well as inside and outside the area it encloses).

In the following, we study   dynamics of the field configuration,
whose properties 
are not affected by a cutoff in momentum space.
Such a requirement is satisfied by  
\be
f(\OM{k})=1,
\label{f11}
\ee
which leads to a rather curious complication.
 Namely, for such a choice 
 \be
\lara{\BD{E}(t,\BD{r})}= -\frac{q\rhat}{2\pi^2}\int_0^\infty
d\OM{k} \partial_r\B{ \frac{\sin(\OM{k}r)}{\OM{k}r}\cos(\VAREPS{k}t)},
 \ee
where $\partial_r=\partial/\partial r$.
The aforementioned 
complication arises from the fact that the above integral is divergent.
However, the interesting thing is that 
if we were allowed to replace 
$\int_0^\infty d\OM{k} \partial_r(\cdots)$ with 
$\partial_r\int_0^\infty d\OM{k}\cdots$,  we would 
have the desired field configuration. 
This suggests the consideration of the 
classical field configuration {\it determined}
by the following formula for the electric field 
\begin{align}
\label{EEE}%
&\BD{E}_\CL(t,\BD{r})= -\rhat\partial_r \phi(t,r),\\
&\phi(t,r)=\frac{q}{2\pi^2r}\int_0^\infty
d\OM{k} \frac{\sin(\OM{k}r)}{\OM{k}}\cos(\VAREPS{k}t),
\label{phi}%
\end{align}
where $\phi(t,r)$ will be referred to as the electric field 
potential.
We have used the word determined in the following sense.
After setting  
\begin{align}
\label{J0}
&{\cal J}_\CL^0(t,\BD{r})= \bnabla\cdot\BD{E}_\CL(t,\BD{r}),\\
&\BD{\cal J}_\CL(t,\BD{r})=-\partial_t \BD{E}_\CL(t,\BD{r}),
\label{JvecC}
\end{align}
 the fields $\BD{E}_\CL$ and $\BD{B}_\CL=\0$ 
satisfy  Proca field equations (\ref{EL}) in the presence of the $4$-current 
$({\cal J}^0_\CL,\BD{\cal J}_\CL)$
\begin{align}
\label{MoMo}
&\bnabla\cdot\BD{E}_\CL(t,\BD{r})={\cal J}^0_\CL(t,\BD{r}), \\
&\bnabla\times\BD{B}_\CL(t,\BD{r})=
\BD{\cal J}_\CL(t,\BD{r})+\partial_t\BD{E}_\CL(t,\BD{r}).
\label{BoBo}
\end{align}
Moreover,  
$\bnabla\times\BD{E}_\CL(t,\BD{r})=-\partial_t\BD{B}_\CL(t,\BD{r})$
and $\bnabla\cdot\BD{B}_\CL(t,\BD{r})=0$,
which  follow  from  the definitions
of $\BD{E}_\CL$ and $\BD{B}_\CL$ in terms of the vector field,
are also satisfied by  the studied field configuration.
Furthermore, local conservation of $4$-current  (\ref{curJ}) 
is preserved by construction   under (\ref{J0}) and (\ref{JvecC}).

The charge  enclosed in the area $\cal V$
will be denoted as $Q_\CL(t,{\cal V})$ and computed 
via
the right-hand side of (\ref{QV}) with  $\lara{\bnabla\cdot\BD{E}(t,\BD{r})}$
replaced by $\bnabla\cdot\BD{E}_\CL(t,\BD{r})$.
The law of periodic charge oscillations 
for the classical field configuration reads
\be
\frac{d^2}{dt^2}Q_\CL(t, \RR^3)=-m^2 Q_\CL(t, \RR^3),
\label{d2Qclass}
\ee
which can be justified similarly as (\ref{d2Q}).

Finally, we comment on the relation of the results presented 
in this section to the ones from \cite{BDPeriodic1}.
It turns out that the  electric field in Sec. 5  of
\cite{BDPeriodic1} can be written as 
$\lara{\BD{E}(t,\BD{r})}=-\rhat\partial_r\hat\phi(t,r)$
and  
\be
\phi(t,r)=-\frac{1}{m^2}\partial_t^2 \hat{\phi}(t,r) \for r\neq t.
\ee
Despite the above mapping, 
there are various  reasons for discussing $\phi(t,r)$
and $\BD{E}_\CL(t,\BD{r})$  in this section.

First, inside  the infinitesimally thin 
spherical shell of the radius $t$,
where quite unusual dynamics occurs, 
the above mapping breaks down and one has to carefully analyze the problem 
from the very beginning. The charge localized 
in such  a shell is actually given by the expression that cannot be anticipated 
from the discussion presented in \cite{BDPeriodic1}.

Second,  explicit expressions for $\phi(t,r)$ and $\BD{E}_\CL(t,\BD{r})$
should be worked out 
because they provide a key ingredient  in the construction of the 
approximate analytical solution discussed in Sec. \ref{Large_sec},
where the counterexample to periodic charge oscillations  
in the quantum Proca theory is discussed.  
Moreover, such expressions are needed for the discussion 
of the  Gibbs-Wilbraham phenomenon in Sec. \ref{Large_sec}.
Furthermore,  it is instructive to contrast 
the cutoff-affected results from Secs. \ref{Sharp_sec} and \ref{Large_sec} with the
 cutoff-free  ones, which is  facilitated by the 
 findings  presented in this  section.

Third,  results discussed in this section provide 
an intriguing  illustration of the phenomenon of periodic charge oscillations, 
making it worthwhile to showcase them explicitly.

\subsection{Electric field potential}
\label{Basic_sub}
The deceptively simple integral determining $\phi(t,r)$ 
encodes  surprisingly rich dynamics. We will now 
discuss 
the results derived in the Appendix, where the 
integral from (\ref{phi}) is examined.

\begin{figure}[t]
\includegraphics[width=\pref\columnwidth,clip=true]{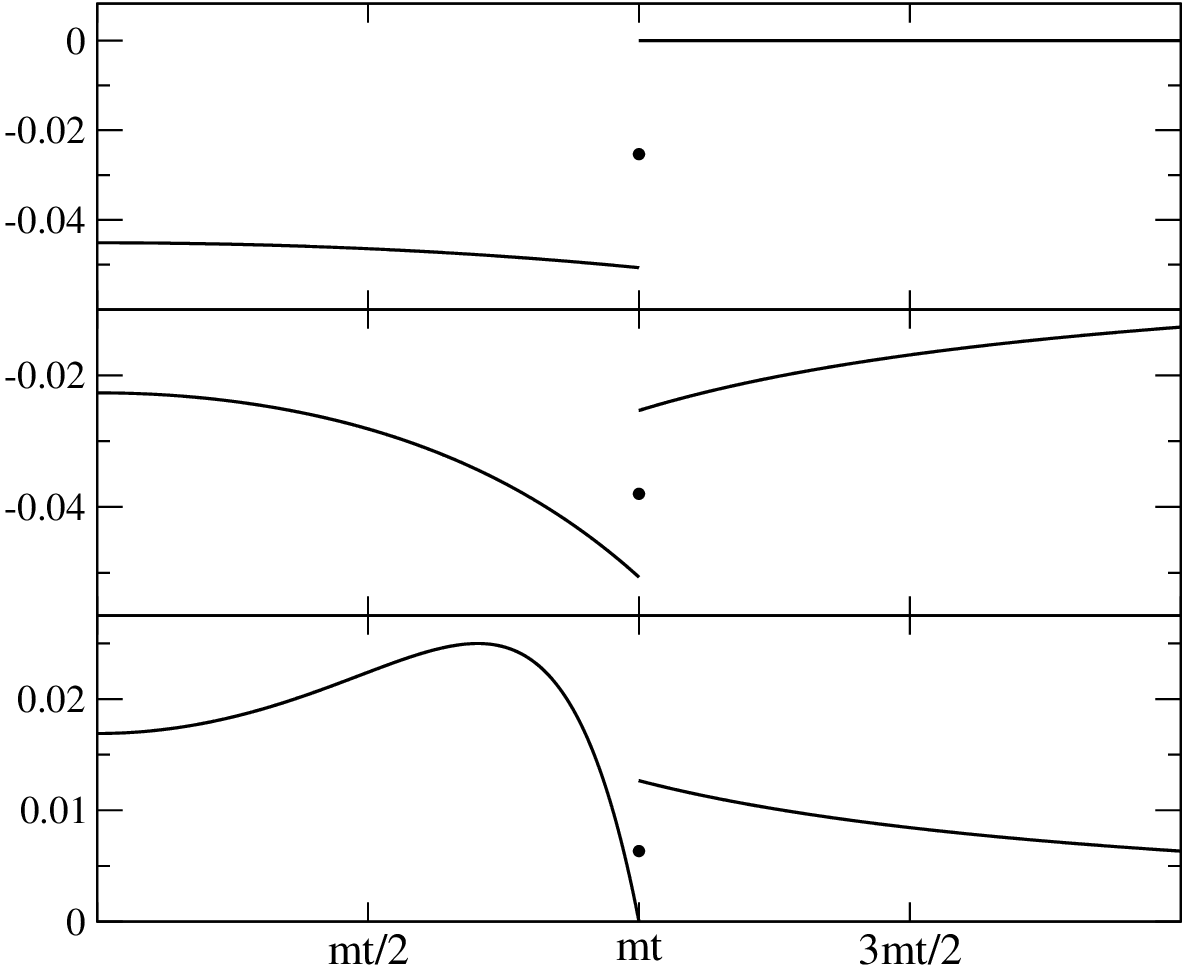}
\caption{$\phi(t,r)\times m^{-1}q^{-1}$, as a function of 
$mr\in[0,2mt]$, computed from (\ref{r_m_tA})--(\ref{r_m_t}).
The panels show results for  $mt=\pi/2,\pi,2\pi$ (top to bottom).
}
\label{phi_trzy}
\end{figure}

We have for  $t=0$ 
\be
\phi(0,r)=\frac{q}{4\pi r},
\label{phit0}
\ee
whereas for $t>0$ we deal with   
\begin{align}
\label{r_m_tA}
&\phi(t,r)=
\frac{q}{4\pi r}\cos( m t) \for r>t,\\
\label{r_m_tB}
&\phi(t,r)=\frac{ q }{4\pi r}\B{\cos( m t)-\frac{1}{2}} \for r=t,\\
&\phi(t,r)=-\frac{ qm t}{4\pi r}\int_0^{r} dx
\frac{J_1\B{ m\sqrt{t^2-x^2}}}{\sqrt{t^2-x^2}} \for 0< r<t,
\label{r_m_t}
\end{align}
where $J_n$  is the Bessel function of the first kind of order $n$.
These formulas are illustrated in Fig. \ref{phi_trzy},
where the results for (\ref{r_m_t}) come from 
the numerical integration performed in \cite{Mathematica141}.
In fact, all our results in this work,  presenting the 
integrals that we cannot analytically  compute,
have been obtained in such a way.

As the above remark indicates, 
we are unaware of the closed-form expression for (\ref{r_m_t}). 
However, we note  that 
the  evaluation of  (\ref{r_m_t}) can be reduced  
to the computation of 
\be
\int_0^{r} dx J_0(m \sqrt{t^2-x^2}) 
\label{J00}
\ee
via  (\ref{dJ0}).
 To the best of our knowledge, 
(\ref{J00}) is analytically known only for  $r=t$, where
it is equal to $\sin( m t)/m$, as  can be inferred from 
(\ref{J0sin}). Such a result can be used for showing that 
\be
\phi(t,r=t^-)=\frac{q  }{4\pi t}\B{\cos( m t)-1}.
\label{rtm}
\ee
By combining (\ref{rtm})   with (\ref{r_m_tA}) and (\ref{r_m_tB}), 
we have obtained  the
following compact expression
\be
\phi(t,r=t^-,t,t^+)=\frac{q}{4\pi t}\B{\cos(mt)-\frac{1}{2}}+\frac{q}{8\pi
t}\sign(r-t),
\label{pttt}
\ee
where $\sign$ is the sign function, which returns $0,\pm1$.
We conclude  that there is 
a shock-wave  front in the electric field potential located at $r=t$.

\begin{figure}[t]
\includegraphics[width=\pref\columnwidth,clip=true]{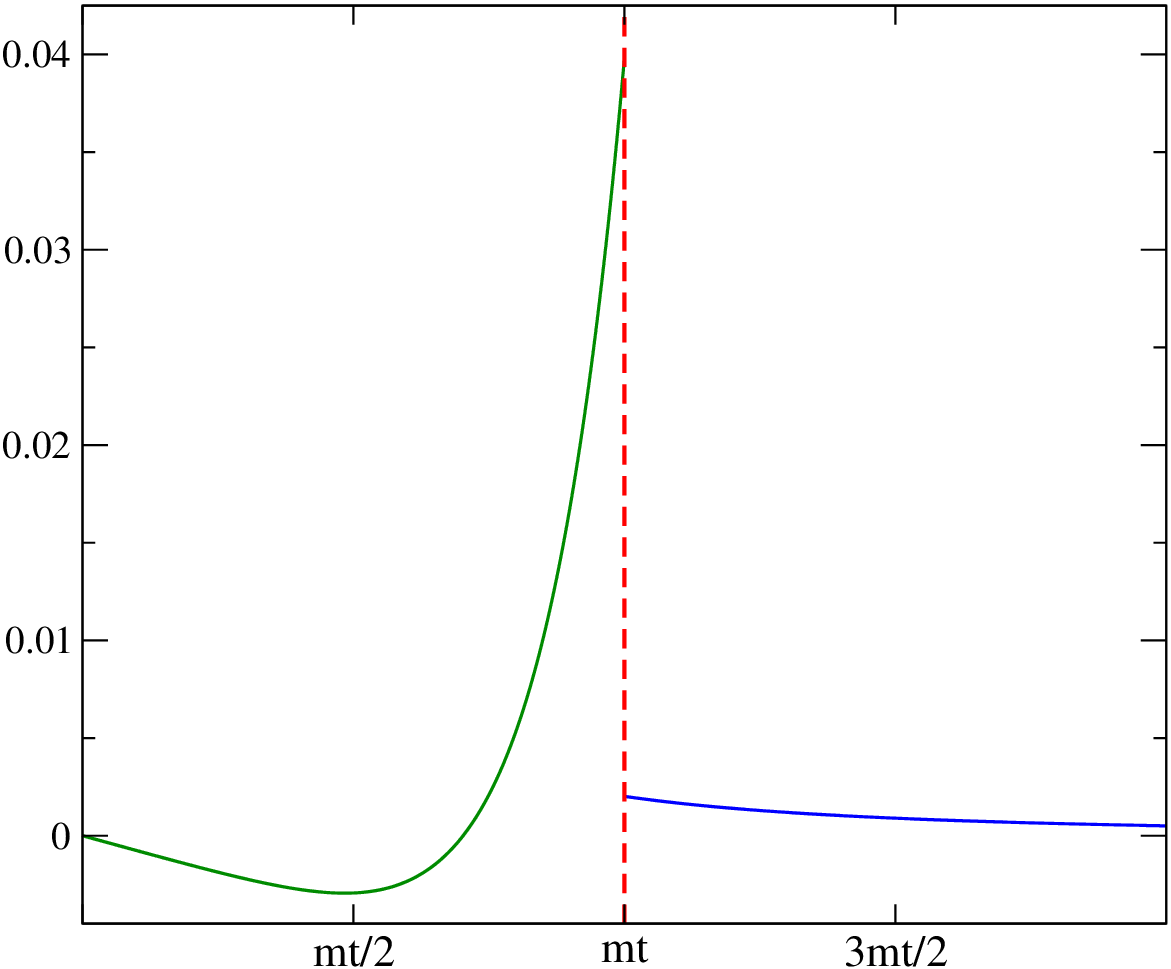}
\caption{$\BD{E}_\CL(t,\BD{r})\cdot\rhat\times m^{-2}q^{-1}$ for $ m t=2\pi$
as a function of $mr\in[0,mt) \cup (mt,2mt]$.
The  green line comes from (\ref{Ein1Ex})
while the blue one  from (\ref{Eout}).
The red    dashed line is meant to indicate that 
we are dealing  with a distributional expression 
 at $r=t$.
This plot is related to the bottom panel
of Fig. \ref{phi_trzy}  via 
$\BD{E}_\CL(t,\BD{r})\cdot\rhat=-\partial_r\phi(t,r)$.
}
\label{EodR}
\end{figure}

\subsection{Electric field}
\label{Electric_sub}
The fact that we have a discontinuity in $\phi(t,r)$
suggests the following organization of the discussion.

{\bf Outside the shock-wave  front.}
For $r>t$, we have  
\be
\BD{E}_\CL(t,\BD{r})=
\frac{q\rhat}{4\pi r^2} \cos( m t),
\label{Eout}
\ee
which  represents the periodically oscillating Coulomb field that  
was introduced  in \cite{BDPeriodic1}.
For  $0 < r<t$, we deal with a considerably
more complicated expression
\be
\BD{E}_\CL(t,\BD{r})=
-\frac{q m t\rhat}{4\pi r^2}\int_0^r
dx\frac{J_1\B{ m\sqrt{t^2-x^2}}}{\sqrt{t^2-x^2}}
+\frac{q m t\rhat}{4\pi r}
\frac{J_1\B{ m\sqrt{t^2-r^2}}}{\sqrt{t^2-r^2}}.
\label{Ein1Ex}
\ee
Given the discussion from Sec. \ref{Basic_sub}, 
we  are  able to simplify the above expression 
 for  $r=t^-$ only. In fact, noting that the first 
 term in (\ref{Ein1Ex}) can be written 
as $\rhat\phi(t,r)/r$,  
 one may easily verify with the help of (\ref{rtm})  that  
\be
\BD{E}_\CL(t,\rhat t^-)=\frac{q\rhat}{4\pi t^2}
\B{\cos(mt)-1+\frac{( m t)^2}{2}}.
\label{deo}
\ee
We have illustrated the above formulas 
in Fig. \ref{EodR}.

{\bf At the shock-wave   front}.
The situation now is more subtle as we are about to
differentiate the discontinuous expression. 
To proceed, we write the electric field potential 
as
\be
\phi(t,r)= 
h(t,r)\theta(r-t) + g(t,r)\theta(t-r),
\label{phiSH}
\ee
where $h(t,r)$ is given by (\ref{r_m_tA}), $g(t,r)$ by 
(\ref{r_m_t}), and the Heaviside step  function is
assumed to satisfy $\theta(0)=1/2$.
Out of such an expression, we get 
\be
\BD{E}_\CL(t,\BD{r})=
-\rhat\partial_r h(t,r)\theta(r-t)
-\rhat\partial_r g(t,r)\theta(t-r)
-\rhat[h(t,r)-g(t,r)]\delta(r-t),
\ee
where the first two terms, after the removal of  Heaviside step  functions,
are given by  (\ref{Eout}) and   (\ref{Ein1Ex}), respectively.
It is  then easy to  show that 
inside  the 
infinitesimally thin  spherical shell of the radius $t$,
i.e. for $t^-<r<t^+$, we have 
\be
\BD{E}_\CL(t,\BD{r}) =
\frac{q\rhat}{4\pi t^2}\cos( m t) -\frac{q\rhat}{4\pi t^2}\B{1-\frac{( m t)^2}{2}}\theta(t-r)
 -\frac{q \rhat}{4\pi t} \delta(r-t).
\label{Edisc}
\ee

\subsection{Charge}
\label{Charge_sub}
Computing the divergence of (\ref{Eout}) and  (\ref{Ein1Ex}), we have obtained
\begin{align}
\label{divE00}
&{\cal J}^0_\CL(t,\BD{r})=0 \for r>t,\\
\label{jedk}
&{\cal J}^0_\CL(t,\BD{r})=\frac{q m t}{4\pi r}\frac{\partial}{\partial r}
\B{\frac{J_1(m\sqrt{t^2-r^2})}{\sqrt{t^2-r^2}}} \for 0 <  r < t
\end{align}
leading to
\begin{align}
\label{Clarge}
&Q_\CL(t,r>t)=0,\\
\label{Crsmall}
&Q_\CL(t,r<t)= q\B{\cos( m t)-1+\frac{( m t)^2}{2}}.
\end{align}
Some   remarks are in order now.

First, from $Q_\CL(t,\mathbb{R}^3)=Q_\CL(t,r>t)+Q_\CL(t,r<t)+Q_\CL(t,t^-<r<t^+)$
and 
\be
Q_\CL(t,\RR^3)=\lim_{r\to\infty}
\int d\BD{S}(\BD{r})\cdot \BD{E}_\CL(t,\BD{r})=q\cos(mt),
\label{QclInfty}
\ee
where $d\BD{S}(\BD{r})$ is the surface element on the sphere of the radius
$r$, we find 
\be
\label{Cshock}
Q_\CL(t,t^-<r<t^+) =q\B{1-\frac{( m t)^2}{2}}.
\ee
We have verified that the same result is obtained via 
the volume integration of the distributional expression 
for $\bnabla\cdot\BD{E}_\CL=-\Delta\phi$ in the 
nearest neighborhood of the shock-wave front. We shall 
not dwell on these (somewhat technical) computations.
Moreover, we observe   that (\ref{QclInfty}) 
obeys law of  periodic charge 
oscillations (\ref{d2Qclass}).

Second, we note that the quadratic time dependence in the above formulas
is valid for all $t>0$ (it does  not come from a
small time expansion). We also note   that (\ref{Crsmall})
can alternatively be obtained from (\ref{deo}) via the surface integral.
In addition, we observe that one may infer  from Fig. \ref{EodR} 
that 
the quadratic in time  growth of (\ref{Crsmall}) is caused by the charge localized 
near the inner edge of the shock-wave  front.
Such a charge is ``neutralized'' by
the charge   that is localized  on the shock-wave  front (\ref{Cshock}).

Third,  local conservation of the $4$-current
implies that the charge cannot be locally created or destroyed.
Still,  it  can escape to spatial infinity or 
emerge   from it. Such a process is possible when 
$\lim_{r\to\infty}\int d\BD{S}(\BD{r})\cdot\BD{\cal J}_\CL(t,\BD{r})\neq0$.
Given the fact that 
$\BD{\cal J}_\CL(t,\BD{r})\propto \rhat$, which can be seen
as the consequence of  the lack of
a distinguished direction in the studied field 
configuration,  $|\BD{\cal J}_\CL(t,\BD{r})|$ vanishing faster 
than $1/r^2$ is needed  for charge conservation.
However,  it follows from 
(\ref{JvecC}) and (\ref{Eout}) that 
$|\BD{\cal J}_\CL(t,\BD{r})|$ is proportional to $1/r^2$ for large-enough $r$.
In fact, for   all $r>t$
\be
\int d\BD{S}(\BD{r})\cdot\BD{\cal J}_\CL(t,\BD{r})=q
m\sin(mt)=-\frac{d}{dt}Q_\CL(t,\RR^3),
\label{DQt}
\ee
and so   there is a charge transport between 
the sphere of the radius $t$  and spatial infinity. 
Such a process is rather peculiar  because 
there is no charge in such a region of space (\ref{divE00}).
This is what we termed as the
empty hose paradox in \cite{BDPeriodic1}. As far as 
we understand it,
such a paradox is resolved in \cite{BDPeriodic1} and so we refer the 
curious reader to this reference.

\subsection{Energy}
\label{Energy_sub}
The energy in the classical Proca theory is 
computed from \cite{Greiner}
\be
\frac{1}{2}\int d^3r (
|\BD{E}_\CL(t,\BD{r})|^2 + |\BD{B}_\CL(t,\BD{r})|^2 
+|\BD{\cal J}_\CL(t,\BD{r})/m|^2 + [{\cal J}^0_\CL(t,\BD{r})/m]^2),
\label{Eproc}
\ee
which is divergent for the studied field configuration.
The infinite value of the energy  raises the question of whether 
the discussed   solution could  be physically relevant besides being 
mathematically interesting 
(as we have argued in Secs. \ref{Basic_sub}--\ref{Charge_sub}) 
and technically useful (as will be shown in Sec. \ref{Large_sec}).
We shall leave open this question but we would like 
to point out that  a   great deal of physically 
reasonable predictions can be 
obtained from  point-charge solutions 
in the Maxwell theory, which describe field configurations 
whose energy is also infinite (see e.g. \cite{MaggioreBook}).
Given the fact that we are about to focus on 
finite-energy field configurations for the 
rest of this work, we shall not dwell on this issue.

\section{Sharp cutoff in  momentum space: general insights}
\label{Sharp_sec}

A quantum field configuration, which serves as a counterexample to the law of periodic
charge oscillations, is constructed in this section. We demonstrate its regularity
(i.e. the absence of shock-wave features) 
and derive exact asymptotic expressions for the electric field, 
the charge density, and the $3$-current.
These expressions are then used to analyze 
the charge localized within a {\it finite region} of space 
and its peculiar dynamics. We show that such a 
charge does not converge to a well-defined  value as the volume of 
the region increases, which is the fundamental reason for the
breakdown of the periodic charge oscillation law.

To proceed, we note that the solution discussed
in Sec. \ref{No_sec}  describes 
the  field configuration whose energy is infinite.
Such an issue 
can be solved by choosing the proper 
function $f$. In our former work
\cite{BDPeriodic1},
we used  {\it smooth} cutoff functions 
$f(\OM{k})=(m/\VAREPS{k})^\gamma$, where $\gamma=2$, $4$, $6$, 
etc. 
This led to the detailed   illustration of the 
phenomenon of periodic charge oscillations. 
In the current work, we introduce the momentum space 
cutoff $\Lambda>0$  via 
\be
f(\OM{k})=\theta(\Lambda-\OM{k}),
\label{fL}
\ee
for    which 
\be
\lara{H}=
q^2\frac{\Lambda}{(2\pi)^2}\BB{1+\frac{1}{3}\B{\frac{\Lambda}{m}}^2}<\infty
\label{tildeH}
\ee
and  rather surprising dynamics takes place.

Equation (\ref{fL}) represents  the sharp cutoff 
function and it leads to 
\begin{align}
\label{tilPhiE}
&\lara{\BD{E}(t,\BD{r})}= -\rhat\partial_r\tilde{\phi}(t,r),\\
&\tilde{\phi}(t,r)=\frac{q}{2\pi^2r}\int_0^\Lambda
d\OM{k} \frac{\sin(\OM{k}r)}{\OM{k}}\cos(\VAREPS{k}t).
\label{tilP}
\end{align}
While discussing expressions of such a sort, we will  invoke   
the sine integral function  rescaled by $2/\pi$,
\be
\Si(x)=\frac{2}{\pi}\int_0^x dx\frac{\sin(x)}{x},
\ee
which has the following asymptotic properties
\cite{OldhamSpanier}
\begin{align}
\label{SiLarge}
&\Si(x\gg1)=1-\frac{2}{\pi} \frac{\cos(x)}{x}-\frac{2}{\pi} \frac{\sin(x)}{x^2}+O(x^{-3}),\\
&\lim_{x\to\pm\infty}\Si(x)=\pm1.
\label{Sinfty}
\end{align}

Below we gather general insights about the studied problem, i.e.
the ones that are valid regardless of the magnitude of $\Lambda$.
While doing so, we highlight the key features of the sharp
cutoff problem.

\subsection{Lack of shock-wave front}
\label{Smothness_sub}

We  note that there are no propagating singularities in 
$\tilde{\phi}(t,r)$ and its derivatives (there is 
no shock-wave front in the sharp cutoff problem).
Given the fact that the smoothness issue has been extensively 
discussed in Sec. \ref{No_sec}, we present below formal 
reasoning proving  the above statement.

To proceed, we write  (\ref{tilP}) as
\begin{align}
&\tilde{\phi}(t,r)= \int_0^\Lambda d\OM{k}\,  {\cal I}(t,r,\OM{k}),\\
&{\cal I}(t,r,\OM{k})= \frac{q}{2\pi^2}
\frac{\sin(\OM{k}r)}{\OM{k}r}\cos(\VAREPS{k}t).
\end{align}
Then, we note that 
${\cal I}:\Omega\times[0,\Lambda]\to\RR$, where 
$\Omega=(0,\infty)\times(0,\infty)$ is an open set in $\RR^2$, 
$(t,r)\in\Omega$,
and 
\be
\frac{\sin(\OM{k}r)}{\OM{k}r}=1-\frac{(\OM{k}r)^2}{3!}+\cdots
\ee
for any $\OM{k}r$. These definitions trigger two
observations.

First, regarding the differentiability class
of ${\cal I}$, we note that 
${\cal I}(t,r,\OM{k})\in C^{\infty}(\Omega)$
for any $\OM{k}\in[0,\Lambda]$.
Second, the map 
\be\Omega\times[0,\Lambda]\ni(t,r,\OM{k})\mapsto
\frac{\partial^\alpha}{\partial t^\alpha}
\frac{\partial^\beta}{\partial r^\beta}{\cal I}(t,r,\OM{k})
\in\RR
\ee
is continuous for all $\alpha, \beta\in\NN_0$.

These  two  observations
imply  that
for $(t,r)\in\Omega$ 
\be
\frac{\partial^\alpha}{\partial t^\alpha}
\frac{\partial^\beta}{\partial r^\beta}
\tilde{\phi}(t,r)= 
\int_0^\Lambda d\OM{k} 
\frac{\partial^\alpha}{\partial t^\alpha}
\frac{\partial^\beta}{\partial r^\beta}
{\cal I}(t,r,\OM{k})
\label{comDER}
\ee
and  
$\tilde{\phi}(t,r)\in C^{\infty}(\Omega)$.
This follows from 
theorem 11.3.3 of  lecture notes 
\cite{JarnickiWyklad_art}, which are particularly well-written
(the  problem of the differentiation of an 
integral is also discussed in standard textbooks).

It should be stressed that  finiteness 
of the energy does not imply smoothness of the 
studied solution.  Indeed, it is shown 
in \cite{BDPeriodic1} that the above-mentioned smooth cutoff 
functions lead to finite-energy solutions that do contain 
a shock-wave singularity. Namely, 
the differentiability class of the
electric field potential in \cite{BDPeriodic1}
is at least $C^{\gamma-2}(\Omega)$
and not more than $C^{\gamma-1}(\Omega)$, most 
likely $C^{\gamma-1}(\Omega)$, and the position of 
the points, where the derivatives of high-enough
order do not exist, changes in time  just 
as in  the problem discussed in Sec.
\ref{No_sec}.
We suspect that  the appearance of the shock-wave front 
is  made possible by  infinite (arbitrarily large)
momenta
involved in the field configuration(s)
studied in Sec. \ref{No_sec}   (Ref. \cite{BDPeriodic1}).

\subsection{Breakdown of law   of periodic charge oscillations}
\label{Lack2_sub}

We will first state the key result of this section and 
then discuss  two ways of proving it.
Namely, we have found that the asymptotic form of the 
electric field is given by the following expression 
\be
\lara{\BD{E}(t,\BD{r})}  = \frac{q\rhat}{4\pi r^2}
\B{\cos(mt)-\frac{2}{\pi}\sin(\Lambda r)\cos(\VAREPS{\Lambda} t)}
+O\B{\frac{\rhat}{r^3}},
\label{flk}
\ee
which via the surface integral calculation yields
\be
\lara{Q(t,r<R)}=4\pi R^2 \lara{\BD{E}(t,\BD{R})}\cdot\hat{\boldsymbol{R}}   = q
\B{\cos(mt)-\frac{2}{\pi}\sin(\Lambda R)\cos(\VAREPS{\Lambda} t)}
+O\B{R^{-1}}.
\label{nkjdf}
\ee
Some remarks are in order now.

First, $\lara{Q(t,r<R)} $ does not converge for $R\to\infty$.
Thereby,
$\lara{Q(t,\RR^3)}$ is undefined and as such 
(\ref{d2Q}) is meaningless (in such a sense 
the law of  periodic charge oscillations
is broken in the studied problem).
Still, there is arguably interesting dynamics exhibited by
$\lara{Q(t,r<R)} $.
Moreover, the spatial dependence of $\lara{Q(t,r<R)} $ is  interesting too.

Second, the first terms in (\ref{flk}) and (\ref{nkjdf})
are the same as in the cutoff-free  problem discussed in Sec. \ref{No_sec}.
The second terms
in these expressions are   cutoff-dependent and their dynamics 
is ``relativistically'' related to  $\Lambda$.
There are two characteristic time scales
 in the discussed problem: 
$2\pi/m$  and $2\pi/\sqrt{m^2+\Lambda^2}$ associated  in the 
Proca theory with $\OM{k}$ equal to $0$ and $\Lambda$ (\ref{Vvecoperator}).

We provide below two derivations of (\ref{flk}).
For this purpose, it is convenient to write the  
electric field as 
\begin{align}
&\lara{\BD{E}(t,\BD{r})}=\frac{q\rhat}{2\pi^2 r^2} I_1
-\frac{q\rhat}{2\pi^2 r} I_2,\\
\label{1I}
&I_1=\int_0^\Lambda d\OM{k}\frac{\sin(\OM{k}r)}{\OM{k}}\cos(\VAREPS{k}t),\\
\label{2I}
&I_2=\int_0^\Lambda d\OM{k} \cos(\OM{k}r)\cos(\VAREPS{k}t).
\end{align}

{\bf Integration by parts approach.} 
We rewrite the first integral  to the form 
\be
I_1=\frac{\pi}{2} \Si(\Lambda r)\cos(mt) + 
\int_0^\Lambda d\OM{k} \sin(\OM{k}r) g_1(\OM{k},t),
\ee
where $g_1(\OM{k},t )= [\cos(\VAREPS{k}t)-\cos(mt)]/\OM{k}$. Integrating 
once by parts, we obtain  
\begin{subequations}
\begin{align}
\label{knjnkj}
&I_1=\frac{\pi}{2} \Si(\Lambda r)\cos(mt) + \delta_1,\\
&\delta_1 = \frac{1}{r} \int_0^\Lambda d\OM{k}
[\cos(\OM{k}r)-\cos(\Lambda r)] g_1'(\OM{k},t),
\end{align}
\label{1ii}%
\end{subequations}
where the prime denotes $\partial/\partial\OM{k}$.
Then, we note that 
\be
\BBB{\delta_1}  \le\frac{2}{r} \int_0^\Lambda d\OM{k} |g_1'(\OM{k},t)| 
= O\B{r^{-1}},
\ee
where  $\int_0^\Lambda d\OM{k} |g_1'(\OM{k},t)|<\infty$.
Proceeding somewhat similarly, but integrating twice by parts now,
we arrive at 
\begin{subequations}
\begin{align}
&I_2=\frac{\sin(\Lambda r)\cos(\VAREPS{\Lambda}t)}{r} + \delta_2,\\
&\delta_2 = \frac{1}{r^2} \int_0^\Lambda d\OM{k}
[\cos(\Lambda r)- \cos(\OM{k}r)] g_2''(\OM{k},t), 
\end{align}
\label{2ii}%
\end{subequations}
where $g_2(\OM{k},t)=\cos(\VAREPS{k}t)$. Next, we observe that 
\be
\BBB{\delta_2}
\le\frac{2}{r^2} \int_0^\Lambda d\OM{k} |g_2''(\OM{k},t)|  
= O\B{r^{-2}},
\ee
where $\int_0^\Lambda d\OM{k} |g_2''(\OM{k},t)|<\infty$.
Finally, we employ (\ref{SiLarge}) to replace $\Si(\Lambda r)$ in 
(\ref{knjnkj}) 
with $1+ O(r^{-1})$ and combine the above results, which leads to 
(\ref{flk}).

{\bf Series expansion approach.} 
We expand $\cos(\VAREPS{k}t)$ into the Maclaurin series 
 and employ the binomial theorem  arriving at 
\be
\cos(\VAREPS{k}t)=\sum_{n=0}^\infty
(-1)^n\frac{t^{2n}}{(2n)!}
\sum_{s=0}^n \binom{n}{s} m^{2(n-s)} \OM{k}^{2s},
\ee
where we do not assume any relation between $m$ and $\OM{k}$
(the Maclaurin expansion of the cosine has infinite convergence radius).
Then, we put such an expression into (\ref{1I}) and (\ref{2I}), 
exchange the order of summation and integration, 
evaluate  the integrals, and sum up the series keeping 
only the leading terms.
As such a procedure straightforwardly 
leads to the expected results, we shall not dwell on 
its  mathematical justification.

To proceed, we observe that 
\be
\int_0^\Lambda d\OM{k} \sin(\OM{k} r)\OM{k}^{2s-1}=
\left\{
\begin{array}{ll}
\frac{\pi}{2}\Si(\Lambda r) & \for s=0\\
O\B{r^{-1}} & \for s\in\ZZ_+
\end{array}
\right.,
\ee
where $O\B{r^{-1}}$ follows from 
formula 2.633.1 of \cite{Ryzhik}. 
As can be easily verified, this leads to 
\be
I_1=\frac{\pi}{2}\Si(\Lambda r)\cos(m t)  + O(r^{-1}).
\label{1i}
\ee
Next, we note that formula 2.633.2 of \cite{Ryzhik}
results in  
\be
\int_0^\Lambda d\OM{k} \cos(\OM{k} r)\OM{k}^{2s}=
\frac{\sin(\Lambda r)}{r} \Lambda^{2s}+O\B{r^{-2}},
\ee
and then one more easy calculation yields
\be
I_2=\frac{\sin(\Lambda r)\cos(\VAREPS{\Lambda}t)}{r} + O\B{r^{-2}}.
\label{2i}
\ee
Just as (\ref{1ii})
and (\ref{2ii}), (\ref{1i}) and (\ref{2i}) lead to (\ref{flk}).

\subsection{Charge dynamics}
\label{Charge2_sub}

We have found the following asymptotic  expression for 
the  charge density 
\be
\lara{{\cal J}^0(t,\BD{r})}= -\frac{q\Lambda}{2\pi^2 r^2}\cos(\Lambda r)\cos(\VAREPS{\Lambda} t)
+ O\B{r^{-3}},
\label{J0exp}
\ee
where the leading order contribution is given by the divergence
of the leading order contribution to (\ref{flk}). 
Such  a property
should not be regarded  as an obvious consequence of
(\ref{j00}), 
which the following example from Sec. \ref{Lack2_sub}
illustrates. Namely, while $\partial_r I_1=I_2$,
the leading order  contribution to $I_2$, $\sin(\Lambda
r)\cos(\VAREPS{\Lambda}t)/r$, is not obtained 
by the action of $\partial_r$ on the leading order contribution
to $I_1$, $\pi\Si(\Lambda r)\cos(m t)/2$. To arrive at the correct 
 leading order  contribution to $I_2$ via differentiation of $I_1$, 
 one also has to take into account    $\partial_r\delta_1$.
To obtain  (\ref{J0exp}), we have written the studied quantity 
as
\begin{align}
&\lara{{\cal J}^0 (t,\BD{r})}=\frac{q}{2\pi^2 r} I_3, \\
&I_3=\int_0^\Lambda d\OM{k} \OM{k} \sin(\OM{k}r) \cos(\VAREPS{k}t),
\end{align}
and  used  the integration by parts approach
from Sec. \ref{Lack2_sub} to derive  the
expression for $I_3$ resulting in (\ref{J0exp}).

We have also found the following asymptotic  expression for 
the $3$-current
\be
\lara{\BD{\cal J}(t,\BD{r})} = \frac{q\rhat}{4\pi r^2}
\B{m\sin(mt)-\frac{2\VAREPS{\Lambda}}{\pi}\sin(\Lambda r)\sin(\VAREPS{\Lambda} t)}
+O\B{\frac{\rhat}{r^{3}}},
\label{Jexp}
\ee
where the leading order contribution is given by the action of $-\partial_t$
on the leading order contribution to (\ref{flk}),
as suggested by (\ref{VpartT}).
However, given the above remarks, 
we have written the quantity of 
interest as 
\begin{align}
&\lara{\BD{\cal J}(t,\BD{r})}= \frac{q\rhat}{2\pi^2 r^2} \tilde{I}_1
-\frac{q\rhat}{2\pi^2 r} \tilde{I}_2,\\
&\tilde{I}_1= \int_0^\Lambda d\OM{k}\frac{\VAREPS{k}}{\OM{k}}\sin(\OM{k}r)\sin(\VAREPS{k}t),\\
&\tilde{I}_2=\int_0^\Lambda d\OM{k} \VAREPS{k}\cos(\OM{k}r)\sin(\VAREPS{k}t),
\end{align}
and  employed  the integration by parts approach
from Sec. \ref{Lack2_sub}
to compute the  expressions for   
$\tilde{I}_1$ and $\tilde{I}_2$ leading to (\ref{Jexp}).
We are ready now to discuss charge dynamics.

To begin, we consider the charge in the ball of the 
radius $R$  (\ref{nkjdf}). 
Its dynamics is caused  by the charge transport  
through the  surface of the ball. Due to $4$-current conservation,
we have 
\be
\frac{\partial}{\partial t}\lara{ Q(t,r<R)}=-4\pi R^2
\lara{\BD{\cal J}(t,\BD{R})}
\cdot\hat{\boldsymbol{R}},
\ee
and so dynamics of $\lara{Q(t,r<R)}$  cannot be suppressed by increasing 
the size of the ball
due to the overall inverse-square decay of $|\lara{\BD{\cal J}(t,\BD{R})}|$ 
(\ref{Jexp}).

Then, we observe that dynamics of   charge density (\ref{J0exp})
is governed by the second term   
in (\ref{Jexp}), $\lara{\BD{\cal J}(t,\BD{r})}_\text{2nd}\propto 
\rhat\sin(\Lambda r)\sin(\VAREPS{\Lambda} t)/r^2$.
To explain this observation, we consider the charge between 
the spheres having  radiuses $r$ and $r+dr$, 
$4\pi r^2 \lara{{\cal J}^0(t,\BD{r})}dr$ for  infinitesimally small
$dr$, and note that 
\be
\frac{\partial}{\partial t}\B{4\pi r^2 \lara{{\cal J}^0(t,\BD{r})}dr}
=
\frac{\partial}{\partial r}\B{-4\pi r^2 \lara{\BD{\cal
J}(t,\BD{r})}_\text{2nd}\cdot\rhat   }dr
\ee
within the order of approximation employed in (\ref{J0exp}).
The above observation raises the question of what is the role of 
the first term  in (\ref{Jexp}), 
 $\lara{\BD{\cal J}(t,\BD{r})}_\text{1st} \propto \rhat\sin(m t)/r^2$.
It turns out that $\lara{\BD{\cal J}(t,\BD{r})}_\text{1st}$
encodes the same physics as 
$\BD{\cal J}_\CL(t,\BD{r})$ discussed in Sec. \ref{Charge_sub}:
the empty hose paradox-based charge transfer process 
 \cite{BDPeriodic1}. 
Thereby,  the first (second) term in (\ref{Jexp}) is responsible for 
 charge dynamics  described by  the first (second) term in
(\ref{nkjdf}).

\subsection{Additional remark}

To better understand the physical mechanism underlying the breakdown
of the periodic charge
oscillations law,  one may  consider the fundamental 
quantity characterizing the Proca theory: 
the expectation value of the vector field operator, 
$\lara{V(t,\BD{r})}=-\lara{ {\cal J}(t,\BD{r}) }/m^2$. 
In the sharp-cutoff setup, $\lara{V(t,\BD{r})}$  exhibits 
a spatially modulated algebraic decay in $r$,
a feature  that persists under the Proca field equations 
governing  dynamics of $\lara{V(t,\BD{r})}$. 
As a result, the charge density, which is proportional
to $\lara{V^0(t,\BD{r})}$, displays long-range spatial oscillations,
ultimately leading to the violation of the oscillation law.

\section{Sharp cutoff in  momentum space: Large cutoff case}
\label{Large_sec}

We work in this section under the condition $\Lambda \gg m$. 
In this regime, we derive approximate analytical expressions for the 
electric field and its potential, extending the asymptotic 
analysis from Sec. \ref{Sharp_sec} to all distances from the center of the 
studied field configuration. This is achieved with the help of the solution
presented in Sec. \ref{No_sec}. 
In the region near  $r = t$, we observe curious oscillations
in the electric field potential, 
which we trace back to the Gibbs–Wilbraham phenomenon.
 Similar features also appear in the electric field itself 
and we  explain them  accordingly.

\subsection{Electric field potential}

\begin{figure}[t]
\includegraphics[width=\pref\columnwidth,clip=true]{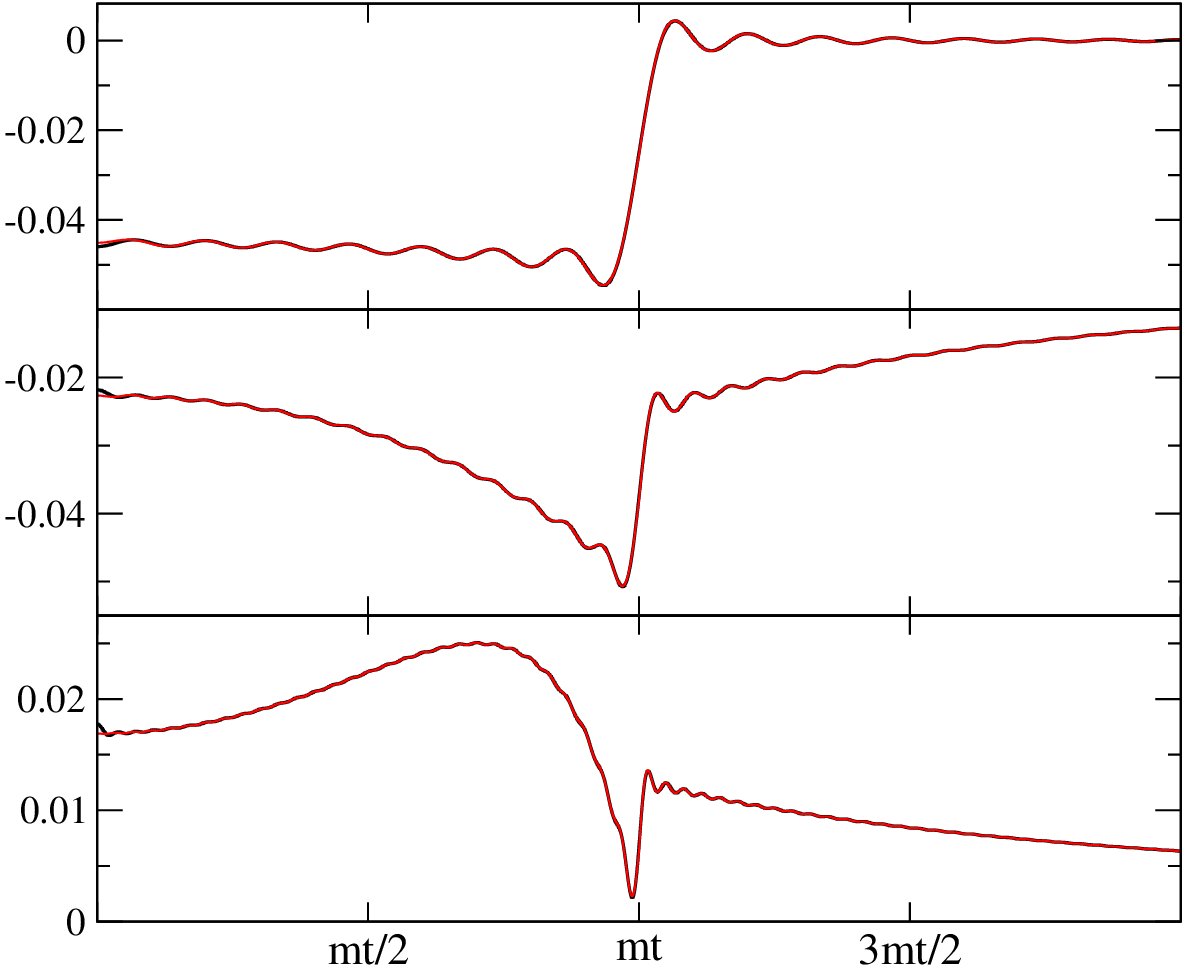}
\caption{$\tilde{\phi}(t,r)\times m^{-1}q^{-1}$ for $\Lambda/m=30$ as a function of 
$mr\in[0,2mt]$. The black lines show  
(\ref{tilP}) while the red ones depict
(\ref{tlPhi}). Both results overlap very well, some differences 
between them are seen near $r=0$ only.
The panels show results for  $mt=\pi/2,\pi,2\pi$ (top to bottom).
For such times (\ref{t_bound}) is satisfied.
}
\label{tilPhi_trzy}
\end{figure}

The dynamics we are now interested  in is depicted in 
Fig. \ref{tilPhi_trzy}, which should be compared to 
Fig. \ref{phi_trzy}. Two  key differences with respect  
to the problem studied in Sec. \ref{No_sec} are the following.
First,  there
is no shock-wave front at $r=t$. Instead of it, there is a rather 
marked  growth of $\tilde{\phi}(t,r)$ around $r=t$. Second, 
there are short-distance oscillations of $\tilde{\phi}(t,r)$.
A more qualitative discussion is provided below.

To proceed, we rewrite (\ref{tilP}) to the form 
\be
\tilde{\phi}(t,r)=\phi(t,r)-
\frac{q}{2\pi^2r}\int_\Lambda^\infty
d\OM{k} \frac{\sin(\OM{k}r)}{\OM{k}}\cos(\VAREPS{k}t).
\label{djk}
\ee
The important thing now is that the employment of 
$\phi(t,r)$, the 
exact solution from Sec. \ref{No_sec} given by (\ref{r_m_tA})--(\ref{r_m_t}), 
confines the integration region 
in (\ref{djk}) 
to high momenta  ($\OM{k}\gg m$).
This observation suggests the following 
approximation
\be
\cos(\VAREPS{k}t)\approx\cos(\OM{k} t),
\label{approx}
\ee
which results  in 
\be
\tilde{\phi}(t,r)\approx \phi(t,r)
-
\frac{q}{8\pi r}
\B{1  
-\Si[\Lambda(r+t)]
+ \sign(r-t)
-\Si[\Lambda(r-t)]
}.
\label{tlPhi}
\ee
Several remarks are in order now.

First, approximation (\ref{approx}) is done under the tacit 
assumption that the time $t$ is small enough. This is seen from
the following  Maclaurin expansion in $m/\OM{k}$
\be
\cos\B{\sqrt{m^2+\OM{k}^2}t}\approx\cos(\OM{k}t) -mt\frac{m}{2\OM{k}}\sin(\OM{k}t).
\label{jden}
\ee
 Indeed, the omission of  
the second term on the right-hand side of (\ref{jden})
requires   $mt(m/\OM{k})/2\ll1$. Given the fact that $\OM{k}\ge\Lambda$
in (\ref{djk}), we arrive at  
\be
 mt \ll 2 \frac{\Lambda}{m}.
 \label{t_bound}
 \ee
It should be kept in mind that (\ref{t_bound}) refers to
the accuracy of (\ref{approx}) and it is 
unclear at the moment how such an approximation 
affects the value of the integral in (\ref{djk}).
We shall not dwell on this technical issue.
Instead, we will compare predictions based on 
(\ref{approx}) to  exact numerical computations
and then quantify  corrections resulting from
the second term 
on the right-hand side of (\ref{jden}).

Second,   (\ref{tlPhi}) 
is compared to (\ref{tilP}) in Fig. \ref{tilPhi_trzy},
where the two expressions  are 
practically indistinguishable, except at the 
smallest
distances from the center of the studied field 
configuration.
The differences between 
these  expressions 
decrease  when   $\Lambda$ increases. 
Moreover, at the risk of stating the obvious, we note that 
approximate result (\ref{tlPhi}) reproduces 
the exact result at  $t=0$
 because  (\ref{approx}) is exactly satisfied 
at such a time instant.

\begin{figure}[t]
\includegraphics[width=\pref\columnwidth,clip=true]{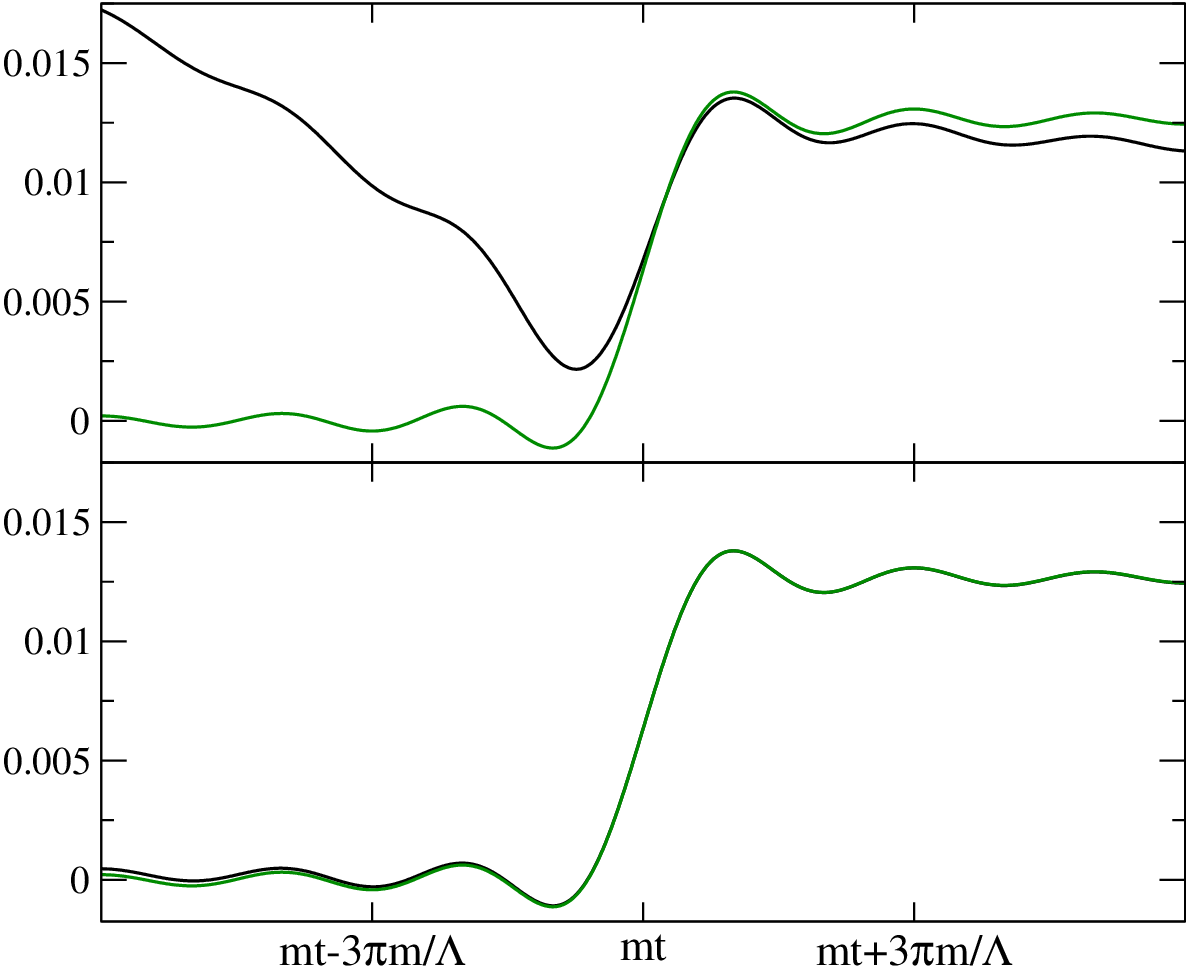}
\caption{$\tilde{\phi}(t,r)\times m^{-1} q^{-1}$ 
 for  $mt=2\pi$ as a function of
 $mr\in[mt - 6\pi m/\Lambda, mt + 6\pi m/\Lambda]$. The upper 
 (lower) panel shows data for $\Lambda/m$ equal to $30$ ($3000$).
The black lines   show (\ref{tilP}) while the green 
ones depict (\ref{koped}).
}
\label{2_larger_L}
\end{figure}

Third, (\ref{tlPhi}) is continuous  at $r=t$ just as the 
exact result.
Namely, the  discontinuity of 
$\phi(t,r)$ is canceled by the discontinuity of the 
expression subtracted from it. 
This  becomes evident upon noting  that the last term 
 of (\ref{pttt}) is exactly  opposite to 
the $\propto\sign(r-t)$ term of (\ref{tlPhi})
evaluated for  $r=t^-,t,t^+$.
In fact, 
(\ref{tlPhi}) yields  
\be
\tilde{\phi}(t,r=t^-,t,t^+)\approx \frac{q}{4\pi t} \B{\cos(mt)-1+\frac{1}{2}\Si(2\Lambda t)}.
\label{ttt}
\ee
Moreover, we propose the following approximation of 
(\ref{tlPhi}) near   $r=t$ 
\be
\tilde{\phi}(t,r)\approx \tilde{\phi}(t,t) 
+ \frac{q}{8\pi t}\Si[\Lambda(r-t)],
\label{koped}
\ee
which has been  obtained by  (i) replacing
$\phi(t,r)$ with
(\ref{pttt}) in (\ref{tlPhi}); (ii) replacing 
$r$ with  $t$ in all expressions in
the resulting formula,  except those being a function of $r-t$.
We note that the accuracy of (\ref{koped}) increases when $\Lambda$ increases,
which is depicted in Fig. \ref{2_larger_L}.
Expression (\ref{koped}) implies that 
 in the region of space  
$|r-t|\le\pi/\Lambda$, 
$\tilde{\phi}(t,r)$ increases roughly by 
\be
\B{1 +2\times 0.08948987}
\frac{q}{4\pi t}
\label{njkwfp}
\ee
because $\Si(x)$  monotonically grows 
from $x=-\pi$ to $x=\pi$ and $\Si(\pm\pi)= \pm1.17897974\cdots$.

\begin{figure}[t]
\includegraphics[width=\pref\columnwidth,clip=true]{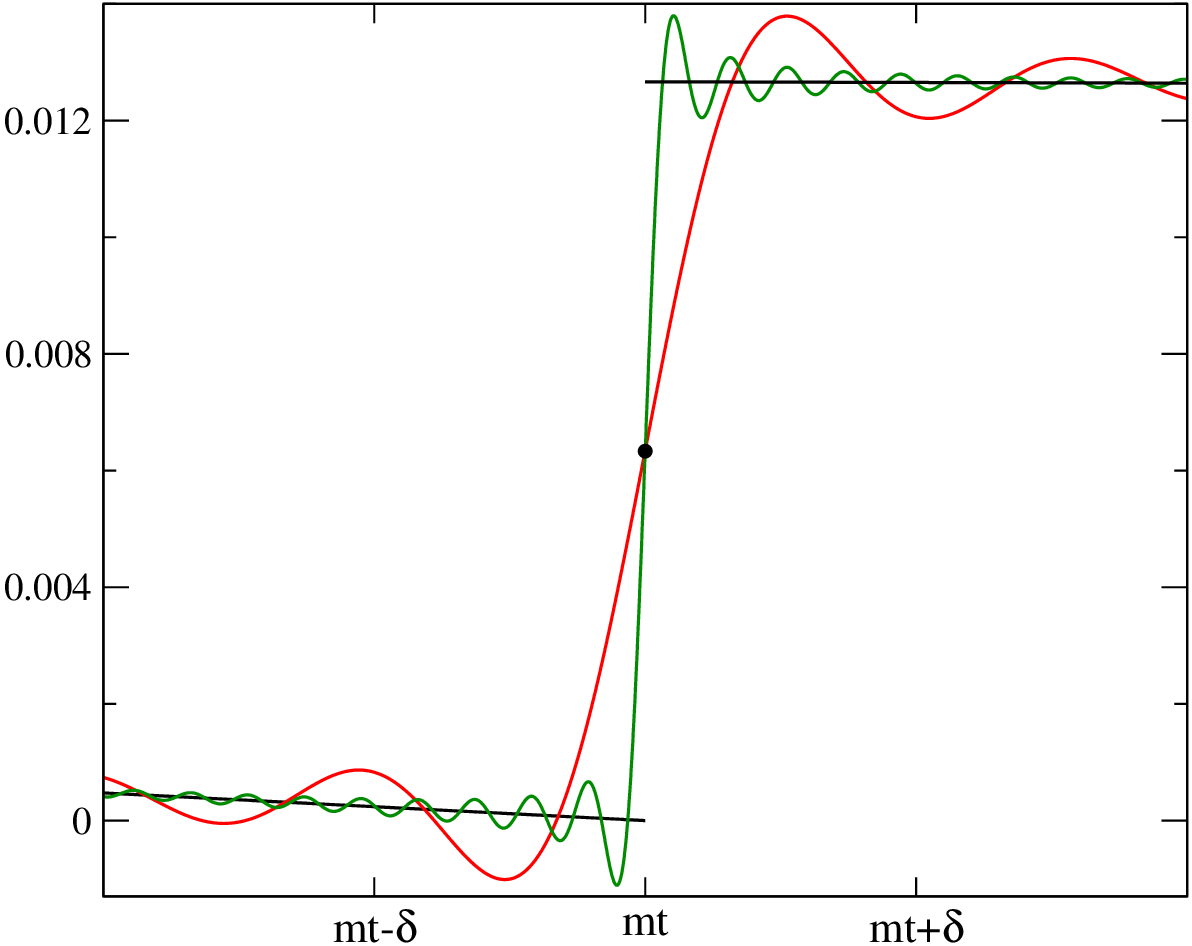}
\caption{The electric field potential for  $mt=2\pi$ as a function of
 $mr\in[mt-2\delta,mt+2\delta]$, where $\delta=0.006$.
The dot and black lines  show $\phi(t,r)\times m^{-1}q^{-1}$.
The red (green)  line
presents $\tilde{\phi}(t,r)\times m^{-1} q^{-1}$
for $\Lambda/m$ equal to $1000$ ($5000$);
(\ref{tilP}) and (\ref{tlPhi}) 
are indistinguishable  on the plotted scale for so large
cutoffs.
The global maximum (minimum) of  the red and green lines 
represents the overshoot (undershoot) that is discussed in 
the context of the Gibbs-Wilbraham phenomenon.
Note essentially the same magnitude  of the overshoot (undershoot)
for large cutoffs used  on this plot. 
}
\label{gibbs}
\end{figure}

Fourth, results (\ref{koped}) and (\ref{njkwfp}) indicate that we 
are dealing here with the Gibbs-Wilbraham phenomenon \cite{HewittGibbs1979}.
Such a phenomenon is traditionally discussed when one computes a Fourier series 
of a  periodic  function having  a jump discontinuity and then tries to reconstruct such
a discontinuity via a truncated Fourier series,  where only   frequencies 
smaller than some cutoff are taken into account.
 It turns out that the reconstructed
function, on one side of the discontinuity,  
overshoots the desired result  by $(\Si(\pi)-1)/2=0.08948987\cdots$ 
times the magnitude  of the jump discontinuity  (this is true in the large cutoff
limit). On the other side of the discontinuity, the truncated Fourier 
series undershoots the desired result by the same amount.
These are  rather counterintuitive features because one would naively 
expect that any overshoot (undershoot) 
should be disappearing  for large cutoffs (see 
\cite{HewittGibbs1979} for a historical sketch of the 
story associated with  this feature).
In our problem, we deal with a Fourier transform
and non-periodic $\phi(t,r)$ representing  the discontinuous function.
Moreover, $\tilde{\phi}(t,r)$ is the truncated expression for 
$\phi(t,r)$ and $\Lambda$  plays the role of the cutoff.
The fact that the Gibbs-Wilbraham phenomenon appears in our problem 
is e.g. seen from (\ref{njkwfp}) if one 
notes that the magnitude of the jump discontinuity of $\phi(t,r)$ is  $q/(4\pi t)$ (\ref{pttt}).
This is further elaborated in Fig. \ref{gibbs}.
Note that   knowledge
of the cutoff-free solution from Sec. \ref{No_sec} is of crucial 
importance in the identification and discussion of the Gibbs-Wilbraham phenomenon
in our studies.

Fifth, approximate result (\ref{tlPhi}) 
 reproduces $\phi(t,r)$ in the large cutoff limit
 in the following sense 
\be
\lim_{\Lambda\to\infty}[\text{right-hand side of (\ref{tlPhi}})]= \phi(t,r),
\label{rhs1}
\ee
where the stress is placed on the {\it pointwise} convergence. Such a result 
follows from  (\ref{Sinfty}) and it does not contradict the above discussion
because it is the lack of {\it uniform}   convergence that is seen in  the 
Gibbs-Wilbraham phenomenon.  However, we would like to stress that 
property (\ref{rhs1}) shall not be taken for granted, 
which will
be evident when we will discuss the  electric field.

The next order improvement over (\ref{approx}) follows
from the consideration of (\ref{jden}).
Under such an approximation,
$\tilde{\phi}(t,r)$ is  equal to the sum of the 
right-hand side  of (\ref{tlPhi}) and 
\be
\delta\tilde{\phi}(t,r)= \frac{qm^2 t}{4\pi^2 r}
\int_\Lambda^\infty d\OM{k} \frac{\sin(\OM{k}r)\sin(\OM{k}t)}{\OM{k}^2}.
\ee
Such an expression can be evaluated via integration 
by parts and elementary manipulations akin to what 
we have employed in  the derivation of (\ref{tlPhi}).
Namely,   
\be
\delta\tilde{\phi}(t,r)=\frac{q m^2 t}{4\pi^2}
\frac{\sin(\Lambda r)\sin(\Lambda t)}{\Lambda  r}
+\frac{q m^2 t(r+t)}{16 \pi r}\B{1 - \Si[\Lambda(r+t)]}
-\frac{q m^2 t(r-t)}{16 \pi r}\B{\sign(r-t) - \Si[\Lambda(r-t)]}.
\label{dndyhje}
\ee

We see from (\ref{dndyhje}) that  
$\lim_{\Lambda\to\infty}\delta\tilde{\phi}(t,r)=0$. Moreover, 
 $\delta\tilde{\phi}(t,r)$ is continuous at $r=t$ and so its 
addition to the
right-hand side  of (\ref{tlPhi}) does not spoil the continuity 
of our analytical approximation for $\tilde\phi(t,r)$.
Finally, regarding the 
differences between the exact  and 
approximate results from 
Fig. \ref{tilPhi_trzy}, we note  that they
 stop being visible  when 
(\ref{dndyhje}) is taken into account.

\subsection{Electric field}
By combining (\ref{tlPhi}) and (\ref{dndyhje}), we have found 
\begin{align}
\lara{\BD{E}(t,\BD{r})}\approx&-\rhat\partial_r\phi(t,r)
+\frac{q\rhat}{4\pi t}\delta(r-t)
-\frac{q\rhat}{8\pi
r^2}\B{1-\frac{(mt)^2}{2}}\B{1-\text{Si}[\Lambda(r+t)]}\nonumber\\
&-\frac{q\rhat}{8\pi r^2} \B{1-\frac{(mt)^2}{2}} 
\B{\sign(r-t)-\text{Si}[\Lambda(r-t)]}-\frac{q\rhat}{4\pi^2 r}
\B{\frac{\sin[\Lambda(r+t)]}{r+t}
+\frac{\sin[\Lambda(r-t)]}{r-t}}\nonumber\\
&+\frac{q m^2 t\rhat}{4\pi^2 r}
\frac{\sin(\Lambda r)\sin(\Lambda t)}{\Lambda r},
\label{wded}
\end{align}
where $-\rhat\partial_r\phi(t,r)$ 
is given by (\ref{Eout}), (\ref{Ein1Ex}), or 
(\ref{Edisc}) depending on the relation between $r$ and $t$.
Several remarks are in order now.

First, by putting   (\ref{Edisc}) and 
$\sign(r-t)=1-2\theta(t-r)$ into (\ref{wded}), we have found
that  (\ref{wded}) is    free from
delta function-like singularities
and continuous at $r=t$.
The same features 
are shared  by 
the exact result for the  electric field, which 
can be inferred from the discussion in Sec. \ref{Smothness_sub}. 
It should be stressed that as far as the  electric field is concerned,
approximation (\ref{approx}) is too crude because it leads
to  $\lara{\BD{E}(t,\BD{r})}$  that is discontinuous at  
 $r=t$  (the $\propto\sign(r-t)$ term  in 
 $\delta\tilde{\phi}$ 
is crucial for ensuring  continuity of 
$\lara{\BD{E}(t,\BD{r})}$  at $r=t$).
Moreover,  (\ref{wded}) leads to 
\begin{align}
\lara{\BD{E}(t,\rhat t)}\approx &-\frac{q\rhat}{4\pi^2 t}\Lambda+
\frac{q\rhat}{4\pi t^2}\cos(mt) 
-\frac{q\rhat}{4\pi t^2}\B{1-\frac{(mt)^2}{2}}
\B{1-\frac{1}{2}\text{Si}(2\Lambda t)}\nonumber\\
&+ \frac{q \rhat}{4\pi^2 t^2}\B{(mt)^2 \frac{\sin^2(\Lambda t)}{\Lambda t}
- \frac{1}{2}\sin(2\Lambda t)}.
\label{edfep}
\end{align}

\begin{figure}[t]
\includegraphics[width=\pref\columnwidth,clip=true]{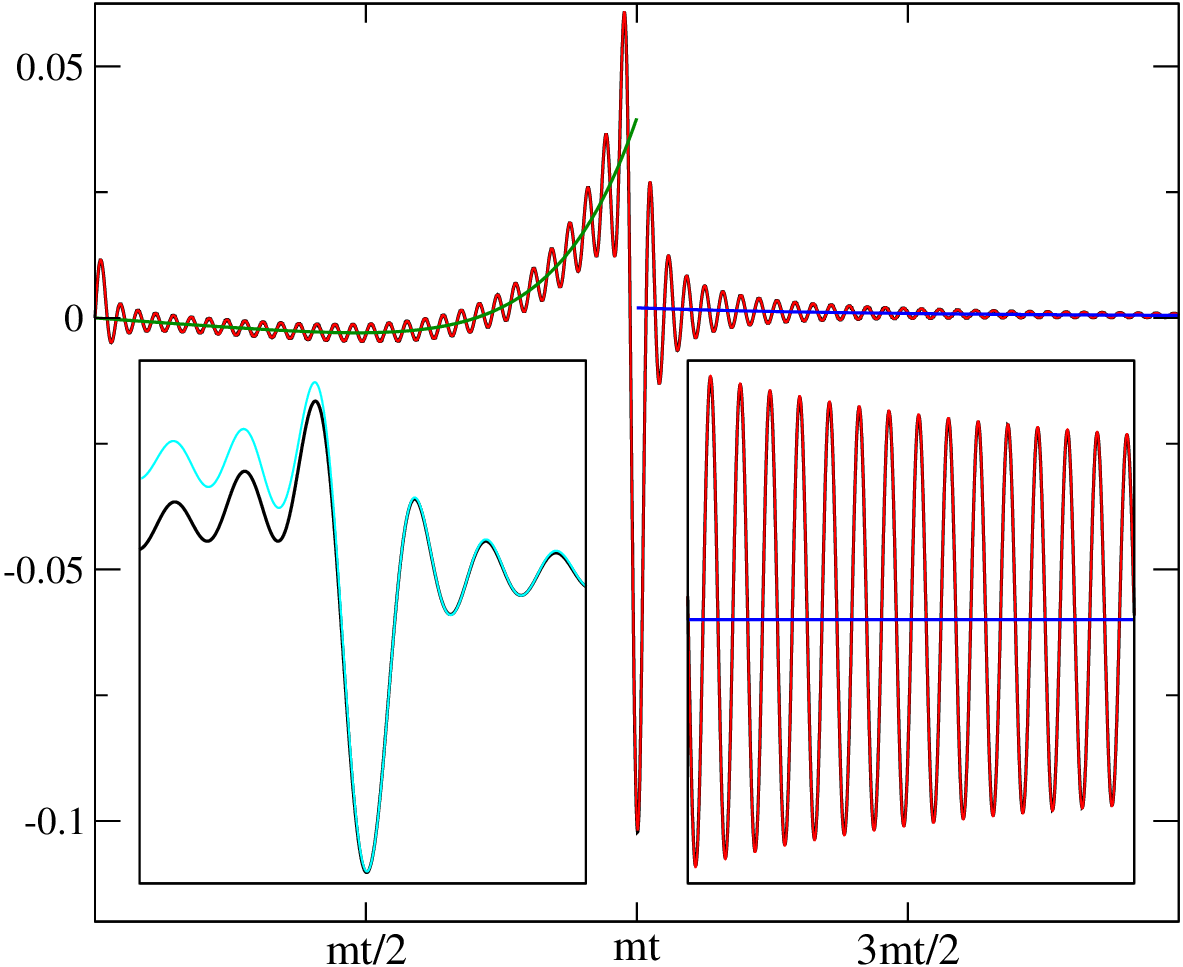}
\caption{$\lara{\BD{E}(t,\BD{r})}\cdot\rhat\times m^{-2}q^{-1}$ for 
$\Lambda/m=30$ and $mt=2\pi$ as a function of 
$mr\in[0,2mt]$. The black line shows (\ref{tilPhiE}) while 
the red one depicts (\ref{wded});
these  lines are practically indistinguishable. 
The green and blue lines represent the cutoff-free  results 
from Fig. \ref{EodR}.
The black lines here and in the bottom panel of 
Fig. \ref{tilPhi_trzy}  are related  via
$\lara{\BD{E}(t,\BD{r})}\cdot\rhat=-\partial_r\tilde{\phi}(t,r)$.
The left inset shows  (\ref{tilPhiE}) and (\ref{123uu}) around $r=t$,
the black (cyan) line represents  the former (latter).
The right inset shows  data, from the 
$mr\in[3mt/2,2mt]$ region of the main plot, multiplied by 
$(mr)^2$. It  illustrates 
the oscillations of the studied solution  around 
the cutoff-free  solution from Sec. \ref{No_sec}
(such a  feature
is not clearly seen in  the main plot for $r>3t/2$).
}
\label{Er_3}
\end{figure}

Second,  (\ref{wded}) is compared to (\ref{tilPhiE})  in 
Fig. \ref{Er_3}, where the 
results obtained  from these expressions  are
practically indistinguishable.
We have also shown 
cutoff-free  results from Sec. \ref{No_sec}  on this figure. The 
solution  obtained  in this section oscillates
around the solution  from Sec. \ref{No_sec}  away from the point, 
where the latter  has the  shock-wave singularity.
Around such a point, however, the two solutions 
have little in common.

Third,  we propose the following approximation of (\ref{wded})
near $r=t$ 
\begin{subequations}
\begin{align}
\label{1uu}
\lara{\BD{E}(t,\BD{r})}&\approx 
\lara{\BD{E}(t,\rhat t)} +\frac{q\rhat}{4\pi^2 t}\Lambda\\
\label{2uu}
&+\frac{q\rhat}{8\pi t^2}\B{1-\frac{(mt)^2}{2}}\Si[\Lambda(r-t)]\\
\label{3uu}
&-\frac{q\rhat}{4\pi t}\delta_\Lambda(r-t),
\end{align}
\label{123uu}%
\end{subequations}
where $\delta_\Lambda(x)=\sin(\Lambda x)/(\pi x)$.
To obtain such a result, we have (i) replaced 
$-\rhat\partial_r\phi(t,r)$ with
(\ref{Edisc}) in (\ref{wded}); (ii) replaced
$r$ with  $t$ in all expressions in 
the resulting formula,  except those being a function of $r-t$.
Formula (\ref{1uu}) oscillates when  $\Lambda\to\infty$
(the divergence of (\ref{edfep}) in such a limit is suppressed 
by the addition of the $\propto\Lambda$ term).
Formula (\ref{2uu}) is given by the product of 
 $[\BD{E}_\CL(t,\rhat t^+)-\BD{E}_\CL(t,\rhat t^-)]/2$
 and $\Si[\Lambda(r-t)]$, where the 
 former is
one-half of the magnitude 
 of the jump of the  cutoff-free
electric field.
Thereby, such an expression gives rise to  the Gibbs-Wilbraham phenomenon
just as (\ref{koped}). Then,  given the fact that 
$\delta_\Lambda(r-t)$ is a nascent delta function, we observe that
formula (\ref{3uu}) is the cutoff-modified 
delta function contribution  from  (\ref{Edisc}).
It provides a dominant contribution to the electric field 
near $r=t$ in the large $\Lambda$ limit
(e.g.  (\ref{3uu}) is responsible for the 
deep global minimum of the electric field in
Fig. \ref{Er_3}). 
Finally, we note that (\ref{123uu}) is compared to  (\ref{tilPhiE})
 in Fig. \ref{Er_3}. This is  done for 
$\Lambda/m=30$ and $mt=2\pi$, where some differences between the two results are seen.
These differences decrease  when $\Lambda$ increases. In particular,
for $\Lambda/m=300$ but still  $mt=2\pi$, 
the two results are practically indistinguishable 
in the region around $r=t$,  which encompasses  several oscillations.

Fourth, we observe that
\be
\lim_{\Lambda\to\infty}[\text{right-hand side
of (\ref{wded})}]\neq -\rhat\partial_r\phi(t,r). 
\label{ldk}
\ee
In fact, such a limit does not exist  
because 
$\propto\sin[\Lambda(r\pm t)]$ terms in (\ref{wded})
do not vanish when 
$\Lambda\to\infty$. Such a situation  
is  a bit surprising because
the studied quantity  is  given by $-\rhat\partial_r\tilde{\phi}(t,r)$
and  our approximate results  
for $\tilde{\phi}(t,r)$ do approach  $\phi(t,r)$
in the pointwise sense   for $\Lambda\to\infty$.
In other words, (\ref{ldk}) suggests that 
the limit of
$\Lambda\to\infty$ 
does not lead to the recovery of the   cutoff-free 
result for the electric field because 
 $[\lim_{\Lambda\to\infty},\partial_r]\tilde{\phi}(t,r)\neq0$.
Such a feature should not be 
seen as the artifact of 
working with  approximate expression (\ref{wded})
for the following reasons.
For one thing, it can be easily exactly shown 
that $[\lim_{\Lambda\to\infty},\partial_r]\tilde{\phi}(0,r)\neq0$.
For another thing,  (\ref{wded})
reproduces exact  (\ref{flk}) in the limit of $r\to\infty$
and    $\propto\sin[\Lambda(r\pm t)]$ terms,
which lead to (\ref{ldk}), 
play a key role in the recovery of such a result.

Fifth, we have simplified  (\ref{wded}) for $r\to\infty$ getting
\be
\lara{\BD{E}(t,\BD{r})} \approx \frac{q\rhat}{4\pi r^2}
\cos(mt)
-\frac{q\rhat}{2\pi^2 r^2}\sin(\Lambda r)
\B{  \cos(\Lambda t)-\frac{m^2t}{2\Lambda}\sin(\Lambda t)  }
+O\B{\frac{\rhat}{r^{3}}}.
\label{flk123456}
\ee
If we now replace the expression in the brackets 
with $\cos(\VAREPS{\Lambda} t)$, which is
 in agreement with  
(\ref{jden}) employed in the derivation of (\ref{wded}),
then  (\ref{flk}) will be recovered.

\section{Summary}
\label{Summary_sec}

We have analyzed  non-equilibrium dynamics of two  long-range 
field configurations in the Proca theory.
The first  has been studied in Sec. \ref{No_sec} while 
the second  in Secs. \ref{Sharp_sec} and \ref{Large_sec}.

These  field configurations  are complementary to each other in the
following sense. 
Namely, we deal with  
the  smooth (discontinuous) 
momentum space decomposition of the electric  field 
potential 
in Sec. \ref{No_sec} (Secs. \ref{Sharp_sec} and \ref{Large_sec}),
which corresponds  to the discontinuous (smooth) 
electric field  potential in real space. 
The complementarity is also seen from the fact that 
the classical (quantum) field configuration 
has been studied in Sec. \ref{No_sec} (Secs.
\ref{Sharp_sec} and \ref{Large_sec}).
Moreover,  the solution from Sec. \ref{No_sec}
is linked  to the one from Secs. \ref{Sharp_sec} and \ref{Large_sec}
via the Gibbs-Wilbraham phenomenon.

The studied field configurations   differ in 
 charge dynamics. While  
the law of periodic charge oscillations 
is obeyed in the problem discussed  in Sec. \ref{No_sec}, 
it is violated in the problem examined in
 Secs. \ref{Sharp_sec} and \ref{Large_sec}.
Since we   overlooked the latter possibility  in
our earlier studies, 
the present work fills an important gap. We have 
re-derived the law of periodic charge
oscillations in Sec. \ref{Charged_sec}  to  explain
why certain field configurations, such as the one studied in 
Secs. \ref{Sharp_sec} and \ref{Large_sec}, do  not obey it.

Then, we note that the discussion in Secs. \ref{No_sec}--\ref{Large_sec} 
has been mainly carried 
out under the assumption that $t>0$. However, one may easily 
extend the obtained results to an arbitrary time by noting that 
the electric field potential is symmetric with respect to 
$t\to-t$, which is seen from  (\ref{phi}) and (\ref{tilP})
that are valid for any $t$. Moreover,
 the  electric field  must exhibit the 
very same symmetry as it is given by the negative gradient 
of such a quantity. By the same token, one may argue that
the charge density is also symmetric with respect to
$t\to-t$, whereas the  $3$-current is antisymmetric. 
This brings us to the perpetual evolution scenario
 proposed in \cite{BDPeriodic1}, where one assumes that the
studied field configurations evolve from $t=-\infty$.

We believe that the results presented in this work provide definite 
insights into dynamics of the infrared sector of the Proca theory,
the topic  poorly explored in the literature.
Given the paradigmatic status of the Proca theory, we hope that our 
findings will spark interest in the exploration of similar topics in 
other theories describing massive neutral particles.

\section*{ACKNOWLEDGMENTS}
The research for this publication has been 
supported by a
grant from the Priority Research Area DigiWorld under the 
Strategic Programme Excellence Initiative at Jagiellonian University.

\appendix*
\section{Integral from (\ref{phi})}

The integral from (\ref{phi})  transforms into 
\be
I(a,b)=\int_0^\infty dx \coth(x)\sin[a\sinh(x)]\cos[b\cosh(x)]
\label{Iab}
\ee
after the following change of
variables
\be
\OM{k}= m\sinh(x), \  m r= a, \  m t= b,
\ee
where  $\sinh$, $\cosh$, and $\coth$ stand for
hyperbolic $\sin$, $\cos$,  and $\cot$, respectively.
We are interested in $a,b\ge0$  below.

It turns out that  $I(a,b)$ was computed  in Appendix 
C of \cite{BDPeriodic1}, along with four other similar integrals,
where  the  differentiability of a certain integral was discussed.
We will now quote  the results presented  there for the sake of 
completeness and
simplify one of them.

Namely, for $a>b>0$
\be
I(a,b)=\frac{\pi}{2}\cos(b),
\label{Iab1}
\ee
whereas for $a=b>0$
\be
I(b,b)=\frac{\pi}{2}\cos(b)-\frac{\pi}{4}.
\label{Ibb}
\ee

Then, for $b>a>0$
\be
I(a,b)=\frac{\pi}{2}\cos(b)-\frac{\pi}{2}
+\frac{b\pi}{2}\int_a^b dx \frac{J_1(\sqrt{b^2-x^2})}{\sqrt{b^2-x^2}},
\ee
which can be simplified by means of 
\be
\frac{d}{dx} J_0(x)=-J_1(x)
\label{dJ0}
\ee 
and the following version of formula 6.677.6 from \cite{Ryzhik}
\be
\int_0^b  dx J_0(\sqrt{b^2-x^2}) = \sin(b).
\label{J0sin}
\ee
Indeed, using the above formulas,    we arrive at 
\be
I(a,b)=-\frac{b\pi}{2}\int_0^a dx \frac{J_1(\sqrt{b^2-x^2})}{\sqrt{b^2-x^2}},
\label{IabJ}
\ee
which  
is  used in the 
main body of this work.

Finally,  it  follows from (\ref{Iab}) that 
$I(0,b)=0$ for $b\ge0$.
Moreover, $I(a,0)=\pi/2$ for $a>0$, 
which is seen from the  momentum space representation of 
$I(a,b)$ given by   (\ref{phi}).


\end{document}